\documentclass[acmsmall]{acmart}
\AtBeginDocument{%
  }

\usepackage{xspace}
\usepackage{graphicx}
\usepackage{booktabs}
\usepackage{tabularx}
\usepackage{subcaption}
\usepackage{algorithm}
\usepackage{algorithmicx}
\usepackage{algpseudocode}
\usepackage{xcolor}
\usepackage[htt]{hyphenat}

\usepackage[most]{tcolorbox}
\usepackage{listings}
\usepackage[table]{xcolor}
\usepackage{enumitem}
\usepackage{multirow}
\usepackage{pifont}
\usepackage{soul}

\definecolor{sqlkeyword}{rgb}{0, 0, 0.8}   
\definecolor{sqlstring}{rgb}{0.8, 0.4, 0}    
\definecolor{sqlcomment}{rgb}{0, 0.5, 0}   

\lstdefinestyle{sqlstyle}{
  language=SQL,
  basicstyle=\ttfamily\scriptsize,
  keywordstyle=\color{sqlkeyword}\bfseries,
  stringstyle=\color{sqlstring},
  commentstyle=\color{sqlcomment}\itshape,
  breaklines=true,
  breakindent=0pt, 
}

\algrenewcommand{\algorithmiccomment}[1]{\bgroup\color{gray}\textit{// #1}\egroup}
\algrenewcommand{\algorithmicrequire}{\textbf{Input:}}
\algrenewcommand{\algorithmicensure}{\textbf{Output:}}
\algnewcommand{\Func}[1]{\mathtt{#1}}

\newcommand{\stab}{\vspace{1.2ex}\noindent}
\newcommand{\stitle}[1]{\stab\noindent{\textbf{#1}}}
\newcommand{\etitle}[1]{\vspace{1mm}\noindent{\underline{\em #1}}}

\newcommand{\vs}{\textit{vs.}\xspace}
\newcommand{\ie}{\textit{i.e.,}\xspace}
\newcommand{\eg}{\textit{e.g.,}\xspace}

\newcommand{\model}{\textsc{DeepEye-SQL}\xspace}

\NewDocumentCommand{\yuyu}{ mO{} }{\textcolor{blue}{\textsuperscript{\textit{Yuyu}}\textsf{\textbf{\small[#1]}}}}

\NewDocumentCommand{\boyan}{ mO{} }{\textcolor{purple}{\textsuperscript{\textit{Boyan}}\textsf{\textbf{\small[#1]}}}}

\definecolor{colorR1}{RGB}{0, 0, 205}     
\definecolor{colorR2}{RGB}{178, 34, 34}   
\definecolor{colorR4}{RGB}{34, 139, 34}   

\newcommand{\revA}[1]{#1}
\newcommand{\revB}[1]{#1}
\newcommand{\revC}[1]{#1}

\setcopyright{none} 
\setcctype{by}
\acmJournal{PACMMOD}
\acmYear{2026}
\acmVolume{4}
\acmNumber{3 (SIGMOD)}
\acmArticle{158}
\acmMonth{6}
\acmDOI{10.1145/3802035}

\begin{document}

\title{DeepEye-SQL: A Software-Engineering-Inspired Text-to-SQL Framework}

\author{Boyan Li}
\orcid{0009-0009-8391-4687}
\affiliation{%
  \institution{The Hong Kong University of Science and Technology (Guangzhou)}
  \city{Guangzhou}
  \country{China}
  }
\email{bli303@connect.hkust-gz.edu.cn}

\author{Chong Chen}
\orcid{0000-0003-1417-2295}
\affiliation{%
  \institution{Huawei Technologies Co., Ltd.}
  \city{Shenzhen}
  \country{China}
  }
\email{chenchong55@huawei.com}

\author{Zhujun Xue}
\orcid{0009-0001-9303-2153}
\affiliation{%
  \institution{Huawei Technologies Co., Ltd.}
  \city{Shenzhen}
  \country{China}
  }
\email{xuezhujun@huawei.com}

\author{Yinan Mei}
\orcid{0009-0001-3841-7149}
\affiliation{%
  \institution{Huawei Technologies Co., Ltd.}
  \city{Shenzhen}
  \country{China}
  }
\email{yinan.mei@huawei.com}

\author{Yuyu Luo}
\orcid{0000-0001-9530-3327}
\affiliation{%
  \institution{The Hong Kong University of Science and Technology (Guangzhou)}
  \city{Guangzhou}
  \country{China}
  }
\email{yuyuluo@hkust-gz.edu.cn}
\authornote{Yuyu Luo is the corresponding author.}

\begin{abstract}
Large language models (LLMs) have advanced Text-to-SQL, yet existing solutions still fall short of system-level reliability. 
The limitation is not merely in individual modules -- \eg schema linking, reasoning, and verification -- but more critically in the lack of structured orchestration that enforces correctness across the entire workflow.
This gap motivates a paradigm shift: treating Text-to-SQL not as free-form language generation but as a software-engineering problem that demands structured, verifiable orchestration.
We present \model, a software-engineering-inspired framework that reframes Text-to-SQL as the development of a small software program, executed through a verifiable process guided by the Software Development Life Cycle (SDLC).
\revA{
\model integrates four synergistic stages: it grounds user intent through robust schema linking, enforcing relational closure; enhances fault tolerance with N-version SQL generation; ensures deterministic verification via a ``Syntax-Logic-Quality'' tool-chain that intercepts errors pre-execution; and introduces confidence-aware selection that leverages execution-guided adjudication to resolve ambiguity beyond simple majority voting.}
\revB{
Leveraging open-source MoE LLMs ($\sim$30B total, $\sim$3B activated parameters) without any fine-tuning, \model achieves 73.5\% execution accuracy on BIRD-Dev, 75.07\% on the official BIRD-Test leaderboard, and 89.8\% on Spider-Test, outperforming state-of-the-art solutions that rely on larger models or extensive training.
This highlights that principled orchestration, rather than LLM scaling alone, is key to achieving system-level reliability in Text-to-SQL.
}
\end{abstract}



\received{October 2025}
\received[revised]{January 2026}
\received[accepted]{February 2026}

\maketitle


{\small
\noindent\textbf{Code Availability:} 
\href{https://github.com/HKUSTDial/DeepEye-SQL}{https://github.com/HKUSTDial/DeepEye-SQL}. \par
}

\section{Introduction}
\label{sec:intro}

\begin{figure}[t!]
    \centering
    \includegraphics[width=\textwidth]{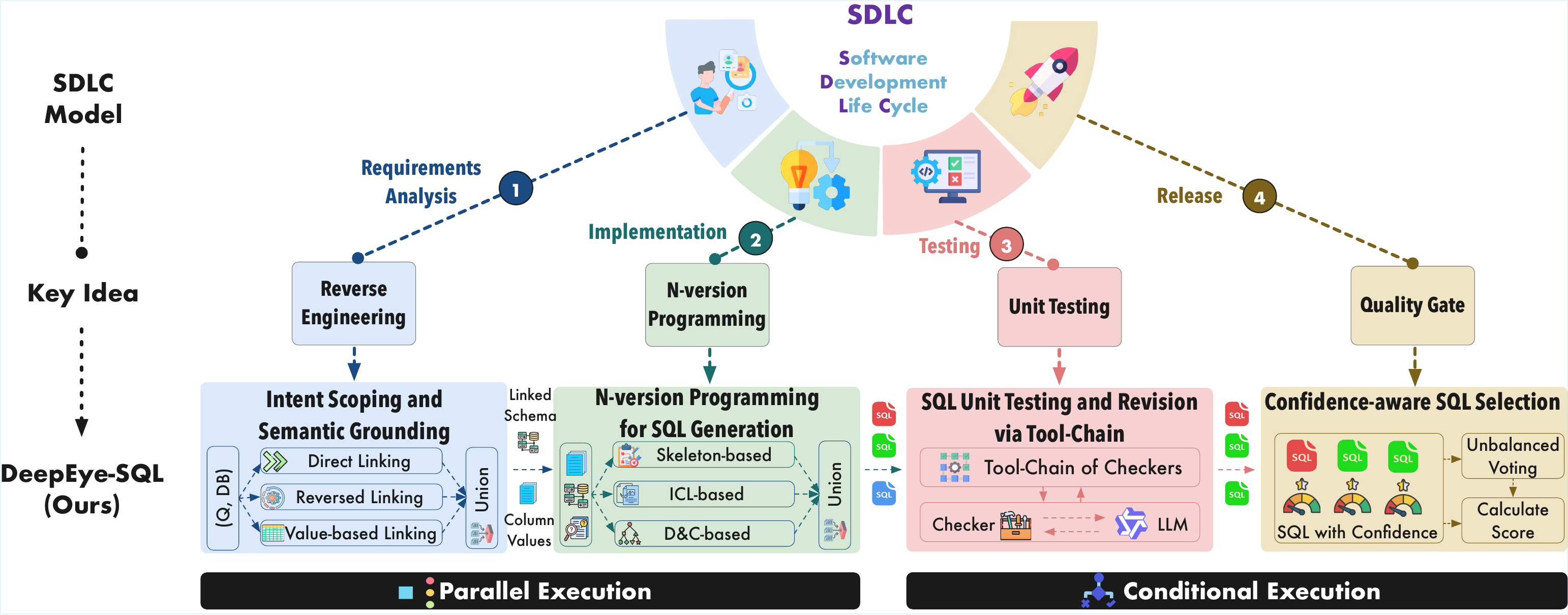}
    \caption{Key Idea of \model.}
    \label{fig:motivation}
\end{figure}

\begin{table*}[t!]
\centering
\caption{\revB{Comparison of \model with representative SOTA Text-to-SQL methods.}}
\label{tab:comparison}
\resizebox{\textwidth}{!}{
\begin{tabular}{l|l|l|l|l}
\toprule
\textbf{Method} & \textbf{Core Paradigm} & \textbf{Verification Mechanism} & \textbf{SQL Selection Strategy} & \textbf{Efficiency \& Cost Profile} \\
\midrule
\multirow{2}{*}{\textbf{CHESS}~\cite{chess}} & \textbf{Linear Inference Workflow} & \textbf{LLM Self-Refinement} & \textbf{Heuristic Majority Voting} & \textbf{High API Cost} \\
 & (Single-path, vulnerable to error propagation) & (Probabilistic, hallucination-prone) & (Standard Self-Consistency) & (Gemini 1.5 Pro, Proprietary) \\
\midrule
\multirow{2}{*}{\textbf{Alpha-SQL}~\cite{alpha-sql}} & \textbf{Search-Centric (MCTS)} & \textbf{Basic Syntax Check} & \textbf{Heuristic Majority Voting} & \textbf{High Inference Latency} \\
 & (Exploration via Monte Carlo Search) & (Ensures executability only) & (Relies on rollout counts) & (MCTS requires massive inferences) \\
\midrule
\multirow{2}{*}{\textbf{OmniSQL}~\cite{omnisql}} & \textbf{Data-Centric SFT} & \textbf{Basic Syntax Check} & \textbf{Heuristic Majority Voting} & \textbf{High Training Cost} \\
 & (Synthetic Data Augmentation) & (Ensures executability only) & (Standard Self-Consistency) & (Requires Full Fine-Tuning) \\
\midrule
\rowcolor{gray!10} \textbf{\model} & \textbf{SE-Lifecycle Centric} & \textbf{Deterministic Unit Testing} & \textbf{Confidence-Aware Selection} & \textbf{Efficient \& Training-Free} \\
\rowcolor{gray!10} \textbf{(Ours)} & \textit{(Fault-Tolerant Design)} & \textit{(Syntax/Logic/Quality Checkers)} & \textit{(Pairwise Adjudication)} & \textit{(\textasciitilde3B Activated Params)} \\
\bottomrule
\end{tabular}
}
\end{table*}

Text-to-SQL is a task that converts natural-language questions into SQL queries over a database~\cite{DBLP:journals/pvldb/LuoLFCT25,nl2sql-survey, deepeye, ncnet, nvbench, nvbench2, DBLP:conf/acl/BaekSHSSGB25, data-agent-survey, elliesql}. 
Large language models (LLMs)~\cite{litecot, deepvis, statqa, aflow, xie2025visjudge, xie2024haichart} have substantially advanced Text-to-SQL, achieving strong results on benchmarks such as Spider~\cite{spider} and BIRD~\cite{bird}.
For example, Alpha-SQL~\cite{alpha-sql} leverages dynamic multi-step reasoning, while XiYan-SQL~\cite{xiyan-sql} improves SQL generation and multi-candidate SQL selection through task-specific fine-tuning.

Despite these advances, state-of-the-art performance on the BIRD dataset remains around 70\% execution accuracy~\cite{bird}, and reliability further degrades in real-world deployments~\cite{supersql,DBLP:journals/pvldb/ChungKGMO25,zhang2021tadoc}. 
This observation indicates that recent advances, while promising, have yet to translate into consistent system-level reliability. 

\textit{The key limitation lies not in the optimization of individual modules -- such as schema linking, reasoning, or post-hoc verification -- \textbf{but in the lack of coherent orchestration that enforces correctness across the entire workflow}}~\cite{xiyan-sql,alpha-sql,din-sql,opensearch-sql,dail-sql,chess,chase-sql}.
As a result, current Text-to-SQL solutions struggle to:
\textit{(i) define what should be built}, which requires precisely determining the semantic scope of the user question and grounding it to the relevant database entities through comprehensive schema linking and value retrieval;  
\textit{(ii) implement the solution (\ie SQL generation)}, which involves generating executable SQL queries that faithfully capture the inferred semantics while maintaining diversity and completeness across complex reasoning paths;  
\textit{(iii) verify its correctness}, which requires systematically validating the structural, logical, and semantic correctness of the generated SQL through interpretable checks~\cite{nl2sql-bugs}; and  
\textit{(iv) release the generated SQL}, which requires quantifying confidence through multi-source evidence and establishing measurable acceptance criteria for determining whether a generated SQL query is reliable enough for output.

This fundamental gap motivates a paradigm shift: \textbf{\textit{Text-to-SQL should be viewed not simply as a language generation task powered by LLMs, but as a software-engineering problem that requires structured orchestration and verifiable correctness}.} From this perspective, generating a correct SQL query resembles developing a small software program: the system must infer user requirements from natural-language questions with respect to the specified database, realize the intended logic through SQL generation, and ensure its correctness and reliability through systematic testing and quality control. 
Inspired by the Software Development Life Cycle (SDLC)~\cite{sdlc} and illustrated in Figure~\ref{fig:motivation}, we structure the Text-to-SQL generation workflow as a unified process that integrates requirement analysis from natural-language questions, SQL generation, verification of generated queries, and final release through quality-gated control.
However, implementing this idea in practice is non-trivial and poses several \textbf{challenges}.

First, Text-to-SQL solutions must \textit{infer} user intent from ambiguous natural language and partially observed database schemas. We term this challenge \textit{ambiguous requirement inference}~(challenge~\textbf{C1}).
In the implementation stage, current methods rely on a single reasoning path driven by one model~\cite{din-sql} and one prompt configuration~\cite{supersql}, \textit{resulting in insufficient fault tolerance}~(\textbf{C2}).
In the verification and validation stage, existing methods depend on probabilistic signals such as self-consistency~\cite{alpha-sql} or partial execution feedback~\cite{DBLP:journals/corr/abs-2505-04671} rather than deterministic oracles to assess generated SQL.
We denote this challenge as \textit{unreliable verification and validation}~(\textbf{C3}).
In the release stage, current methods lack \textit{calibrated confidence estimation and measurable acceptance criteria} for evaluating when a generated SQL query is reliable enough for output~(\textbf{C4}).

\begin{figure}[t!]
    \centering
    \includegraphics[width=0.7\columnwidth]{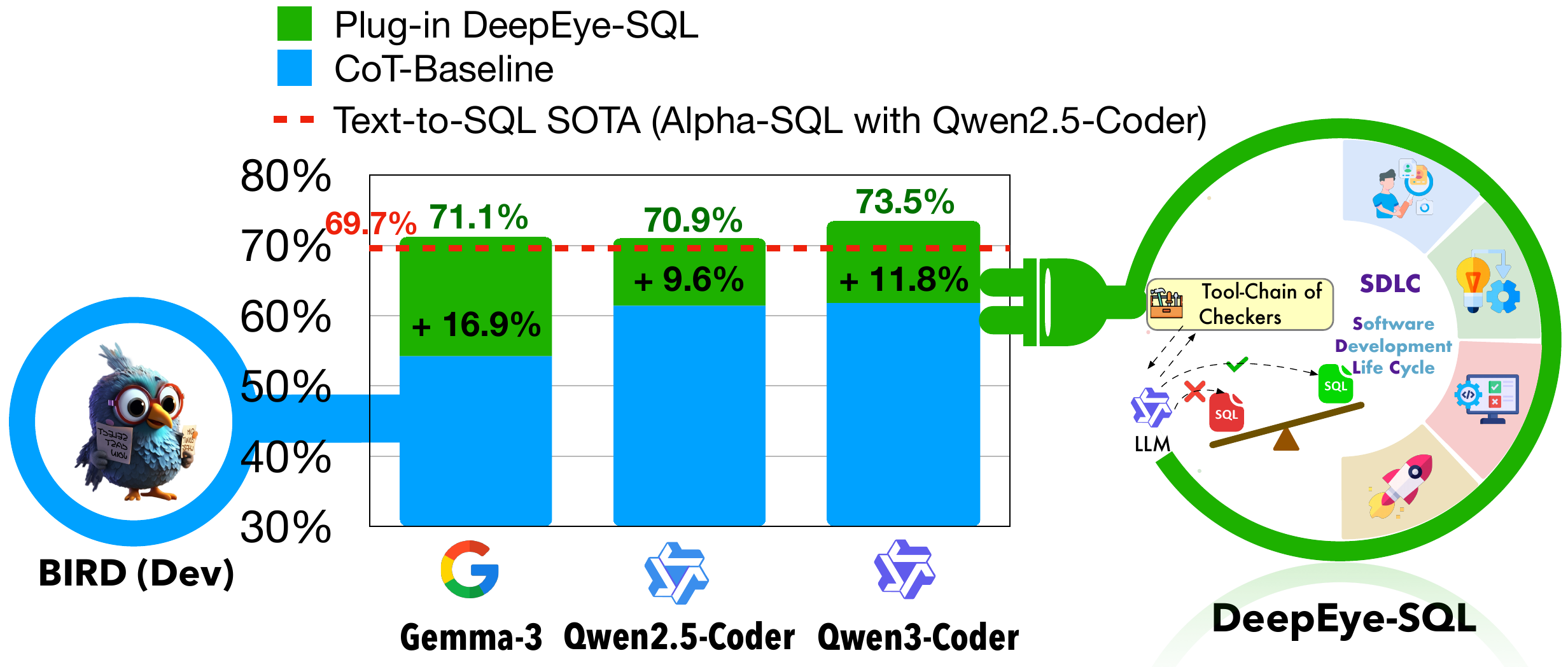}
    \caption{As a plug-and-play framework, \model consistently outperforms SOTA baselines using $\sim$30B LLMs.}
    \label{fig:performance}
\end{figure}

\stitle{Our Methodology.}
To systematically address these challenges, we propose \model, which reframes Text-to-SQL as a verifiable SDLC-style workflow with four stages (Figure~\ref{fig:motivation}).
\revA{In requirements analysis, we propose Semantic Value Retrieval and a fault-tolerant Robust Schema Linking strategy to build a complete, database-grounded specification (addressing \textbf{C1}).}
%
\revA{In implementation, we adopt N-version programming~\cite{n-version-programming} by executing multiple independent generators with diverse reasoning paradigms in parallel to diversify reasoning under a fixed budget, providing fault tolerance (addressing \textbf{C2}).}
%
\revA{For verification and validation, we replace probabilistic self-judgment with a suite of specialized, deterministic checkers that trigger targeted LLM repair, ensuring verifiable correctness (addressing \textbf{C3}).}
%
\revA{Finally, in release, we introduce \revA{Confidence-aware SQL Selection} that dynamically adapts the review process based on execution consensus, yielding a calibrated output (addressing \textbf{C4}).}
As shown in Figure~\ref{fig:performance}, this design enables \model to integrate with diverse LLMs and achieve notable accuracy improvements without any fine-tuning.
\revB{To explicitly distinguish our contribution from existing approaches, Table~\ref{tab:comparison} contrasts \model with representative state-of-the-art methods. Unlike linear workflows (e.g., CHESS~\cite{chess}) or search-centric methods (e.g., Alpha-SQL~\cite{alpha-sql}) that rely heavily on single-path reasoning or probabilistic self-correction, \model enforces a \textit{deterministic SE-Lifecycle} paradigm. We also incorporate a fault-tolerant design via \textit{N-Version Programming} to mitigate single-point reasoning failures. Furthermore, by replacing stochastic checks with a rigorous \textit{Unit Testing Tool-Chain} and introducing \textit{Confidence-Aware Selection}, we systematically address the inherent unreliability of LLM self-reflection.}




\stitle{Contributions.}
This paper makes the following contributions:


\stab\revB{(1) \textbf{A Novel SE-Inspired Framework.} We propose \model, which reframes Text-to-SQL as a verifiable SDLC workflow. Unlike prior frameworks constrained by \textit{fragile, linear inference chains} where errors propagate unchecked, \model enforces a \textit{robust engineering lifecycle} (Requirement $\to$ Implementation $\to$ Testing $\to$ Release) to guarantee system-level reliability.}

\stab\revB{(2) \textbf{N-Version Programming for Diversity.} Departing from standard self-consistency methods that rely on stochastic sampling (same prompt, high temperature), we introduce an \textit{N-Version Programming} paradigm. By orchestrating independent generators with distinct reasoning strategies, we achieve true algorithmic fault tolerance and broader coverage of complex queries.}
    
\stab\revB{(3) \textbf{Deterministic Verification via Tool-Chain.} To address the confirmation bias inherent in LLM self-correction, we propose a novel \textit{SQL Unit Testing Tool-Chain}. This module replaces probabilistic LLM self-reflection with deterministic, external execution checks (Syntax, Logic, and Data Quality), providing precise feedback that grounds the revision process.}

\stab\revB{
(4) \textbf{Extensive Experiments} demonstrate that \model achieves state-of-the-art performance on challenging benchmarks. Leveraging open-source MoE LLMs with only $\sim$3B activated parameters, \model attains an execution accuracy of 73.5\% on BIRD-Dev, 75.07\% on the held-out BIRD-Test set, and 89.8\% on Spider-Test, outperforming larger or fine-tuned baselines.
}

\section{DeepEye-SQL Overview}
\label{sec:overview}


\begin{figure*}[t!]
    \centering
    \includegraphics[width=\linewidth]{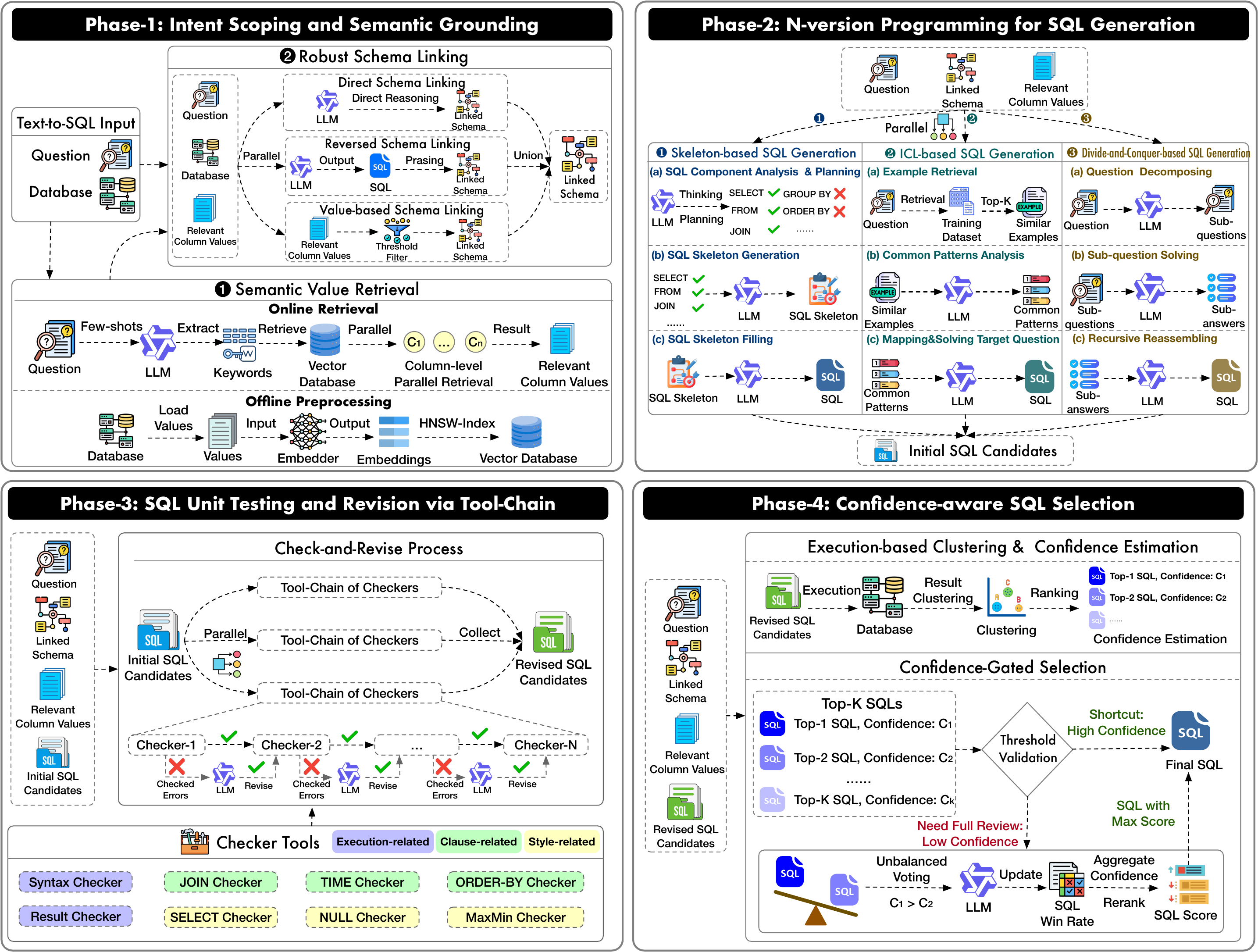}
    \caption{\model Overview.}
    \label{fig:overview}
\end{figure*}

\model is systematically organized as a multi-stage pipeline inspired by the core principles of the Software Development Life Cycle (SDLC). Rather than reproducing the entire SDLC process literally, it abstracts and automates its essential workflow for Text-to-SQL. Conceptually, \model serves as a compressed and self-contained microcosm of software development, where the ``software'' being constructed is a single, correct SQL query. Following this paradigm, the framework comprises four stages (Figure~\ref{fig:overview}), each aligned with a corresponding phase of the SDLC.


\stitle{Phase-1: Intent Scoping and Semantic Grounding.}
Mirroring the initial phase of any engineering project—answering ``What should be built?''—this stage is dedicated to accurately interpreting user intent and defining the problem's scope. It employs \textit{Semantic Value Retrieval} to ground the query in the database's actual data and \textit{Robust Schema Linking} to identify necessary tables and columns. 
\revA{This linking module uses a fault-tolerant hybrid strategy that integrates complementary linking paradigms to avoid the ``\textit{single point of failure}'' problem~\cite{spof}, ensuring a comprehensive and fault-tolerant specification.}


\stitle{Phase-2: N-version Programming for SQL Generation.}
Analogous to the implementation phase, this stage generates SQL queries. To enhance robustness, we employ a strategy inspired by \textit{N-version programming}~\cite{n-version-programming}, a fault-tolerance technique where multiple, independent implementations are created for the same problem.
%
\revA{Our framework instantiates this by producing a diverse set of SQL candidates in parallel from three distinct generators, each utilizing a unique reasoning paradigm.}
%
\revA{It is crucial to distinguish this approach from test-time scaling techniques like self-consistency~\cite{self-consistency}, which generate diversity by sampling multiple outputs from a single generator. In contrast, our method achieves a more principled and profound diversity, akin to true N-version programming, as each generator employs a fundamentally different reasoning process.} This engineered diversity ensures a broader exploration of the solution space, significantly increasing the probability of producing at least one correct query, especially for complex scenarios where a single reasoning path might fail.


\stitle{Phase-3: SQL Unit Testing and Revision via Tool-Chain.}
This phase embodies the critical software engineering principle of \textit{Unit Testing}~\cite{unit-testing}. Its purpose is to systematically verify the correctness of each generated SQL candidate and revise any found defects. To overcome the known unreliability of LLM self-correction~\cite{self-correction-bias}, our framework externalizes this process, emulating a rigorous, automated testing loop.
\revA{Each SQL candidate is passed through a \textit{Tool-Chain of Checkers}---a suite of specialized, deterministic tools where each checker acts as a test case for a specific unit of functionality. These checkers systematically validate structural and semantic constraints, and if a flaw is detected, provide \textit{an explicit and actionable directive} to the LLM for a targeted revision, mirroring a formal bug report and fix cycle.}

\stitle{Phase-4: Confidence-aware SQL Selection.}
The final phase corresponds to the release stage, governed by a \textit{Quality Gate}. Instead of simply choosing the most common answer, this stage arbitrates which candidate is reliable enough to be ``released''. Our \textit{Confidence-aware SQL Selection} mechanism performs this task. 
\revA{It estimates the reliability of candidates based on execution consensus and evaluates this against a predefined acceptance criterion. This validation acts as a gateway, determining whether to release the result directly or trigger a deeper, comparative review to ensure the most reliable query is selected.}
\section{The Design Details of \model}
\label{sec:design}


\begin{algorithm}[t!]
\caption{Online Semantic Value Retrieval}
\label{alg:value_retrieval}
\begin{algorithmic}[1]
\small
\Require User Question $Q$, Set of Column Vector Indices $\{\mathcal{I}_j\}_{j=1}^M$, Top-K parameter $K$
\Ensure  Retrieved Values Map $\mathcal{M}_{retrieved}$: $C_j \to \mathcal{V}_j$

\Statex \Comment{Step 1: Extract Keywords from the user question}
\State $\mathcal{K} \leftarrow \Func{LLM-ExtractKeywords}(Q)$

\Statex \Comment{Step 2: Retrieve values in parallel for each indexed column}
\State Initialize $\mathcal{M}_{retrieved} \leftarrow \emptyset$
\ForAll{index $\mathcal{I}_j$ for column $C_j$ \textbf{in parallel}}
    \State $\mathcal{V}_{candidates} \leftarrow \emptyset$
    \ForAll{keyword $k_i \in \mathcal{K}$}
        \State $\mathbf{e}_{k_i} \leftarrow \Func{Embed}(k_i)$
        \State $\mathcal{V}_{partial} \leftarrow \Func{SearchIndex}(\mathcal{I}_j, \mathbf{e}_{k_i}, K)$
        \State $\mathcal{V}_{candidates} \leftarrow \mathcal{V}_{candidates} \cup \mathcal{V}_{partial}$
    \EndFor
    
    \State $\mathcal{V}_{sorted} \leftarrow \Func{SortBySimilarity}(\mathcal{V}_{candidates})$ \Comment{Aggregate and select top-K unique values}
    \State $\mathcal{V}_j \leftarrow \Func{GetUniqueTopK}(\mathcal{V}_{sorted}, K)$
    \State $\mathcal{M}_{retrieved}[C_j] \leftarrow \mathcal{V}_j$
\EndFor

\State \textbf{return} $\mathcal{M}_{retrieved}$
\end{algorithmic}
\end{algorithm}

\subsection{Semantic Value Retrieval}
\label{subsec:value-retrieval}
A key Text-to-SQL challenge is the grounding problem: LLMs lack awareness of specific database values, often generating SQL with hallucinated or mismatched filter values (\eg using \texttt{country = `USA'} when the database stores \texttt{`United States'}). This issue is particularly prevalent for high-cardinality, free-form text columns. Addressing this mirrors the software engineering principle of \textit{dependency resolution}~\cite{dependency-resolution}, which requires a system to be aware of valid data constants it can operate on. To mitigate this, \model incorporates a \textit{Semantic Value Retrieval} module that proactively supplies the LLM with a contextually relevant subset of database values, anchoring the generation process to the ground-truth data.
Our value retrieval process is divided into two distinct stages: an efficient, one-time offline preprocessing stage for index construction, and a rapid online retrieval stage executed at query time.

\stitle{Offline Preprocessing.}
The primary objective of the offline phase is to preprocess and index database values to enable efficient semantic search. This process is executed once for any new database and involves three steps.

\etitle{Selective Value Extraction.} Instead of a brute-force approach that indexes every value---which would be computationally prohibitive and introduce significant noise---we perform selective extraction. We specifically target columns of type~\texttt{TEXT}, as they are the primary source of ambiguity and value-related hallucinations. To further refine this process, we apply heuristics to exclude columns that, despite being text-based, are unlikely to be used in semantic comparisons, such as columns containing UUIDs or exclusively numerical identifiers. This strategic selection minimizes the indexing overhead while enhancing the semantic relevance of the retrieved values.

\etitle{Value Embedding.} For each selected column~$C_j$, we extract its unique values. Each distinct value~$v$~is then encoded into a high-dimensional vector representation~$\mathbf{e}_v$ using a pretrained sentence embedding model, specifically~\texttt{Qwen3-Embedding-0.6B}~\cite{qwen3-embedding}. This embedding transforms discrete text strings into continuous semantic vectors, where values with similar meanings are located closer to each other in the vector space.

\etitle{Vector Indexing.} To facilitate fast similarity search, the generated value embeddings for each column are used to build a vector index. We employ Chroma~\cite{chroma} with the Hierarchical Navigable Small World (HNSW) algorithm~\cite{hnsw} for this purpose. HNSW is highly efficient for approximate nearest neighbor (ANN) search, making it ideal for real-time applications. The result is a set of persistent, per-column vector indices~$\{\mathcal{I}_1, \mathcal{I}_2, ..., \mathcal{I}_M\}$, where each~$\mathcal{I}_j$~is an indexed collection of all value embeddings for a column~$C_j$.

\stitle{Online Retrieval.}
During the online phase, when a user question~$Q$~is received, the system retrieves the most relevant values for that question from the pre-built indices. This process is detailed in Algorithm~\ref{alg:value_retrieval}.

\etitle{Keyword Extraction.}
First, we leverage the LLM to identify potential entities and filter values within the user's question~$Q$. Using a few-shot prompting strategy, we instruct the LLM to extract a set of key terms~$\mathcal{K} = \{k_1, k_2, ..., k_N\}$ that are likely to appear in a~\texttt{WHERE}~clause. For example, from the question ``Show me all papers by authors from France'', the LLM would extract keywords like ``France''.

\etitle{Parallel Multi-Column Retrieval.}
With the extracted keywords, we perform a parallel search across all indexed columns. For each indexed column~$C_j$, we perform the following steps:
\begin{enumerate}[leftmargin=*, topsep=0pt, itemsep=0pt, parsep=0pt]
    \item For every keyword~$k_i \in \mathcal{K}$, we generate its embedding~$\mathbf{e}_{k_i}$~using the same model from the offline phase.
    \item We query the corresponding column index~$\mathcal{I}_j$~with~$\mathbf{e}_{k_i}$~to retrieve the top-$K$ most similar values along with their similarity scores (\eg cosine similarity).
    \item After querying for all~$N$~keywords, we obtain~$N \times K$~candidate values for column~$C_j$. We aggregate these candidates, sort them globally by their similarity scores in descending order, and select the top-$K$ unique values. This yields the final retrieved value set~$\mathcal{V}_j$~for column~$C_j$.
\end{enumerate}


\subsection{Robust Schema Linking}

\label{subsec:schema_linking}

\begin{algorithm}[t!]
\caption{Robust Schema Linking}
\label{alg:schema_linking}
\begin{algorithmic}[1]
\small
\Require User Question $Q$, Database Schema $\mathcal{D}$, Retrieved Values Map $\mathcal{M}_{retrieved}$, Threshold $\theta_{val}$
\Ensure  Final Linked Schema $\mathcal{D}_{linked}$

\Statex \Comment{Step 1: Execute the three linking strategies in parallel}
\State \textbf{begin parallel}
\State \quad $\mathcal{D}_{direct} \leftarrow \Func{LLM\_DirectLink}(Q, \mathcal{D}, \mathcal{M}_{retrieved})$
\State \quad $\mathcal{S}' \leftarrow \Func{LLM\_GenerateSQL}(Q, \mathcal{D}, \mathcal{M}_{retrieved})$
\State \quad $\mathcal{D}_{reversed} \leftarrow \Func{ParseSchema}(\mathcal{S}')$
\State \quad $\mathcal{D}_{value} \leftarrow \Func{FindValueBasedSchema}(\mathcal{M}_{retrieved}, \theta_{val})$
\State \textbf{end parallel}

\Statex \Comment{Step 2: Aggregate the results by taking the union}
\State $\mathcal{D}_{union} \leftarrow \mathcal{D}_{direct} \cup \mathcal{D}_{reversed} \cup \mathcal{D}_{value}$

\Statex \Comment{Step 3: Enforce relational closure to ensure schema connectivity}
\State $\mathcal{D}_{linked} \leftarrow \Func{EnforceClosure}(\mathcal{D}_{union}, \mathcal{D})$

\State \textbf{return} $\mathcal{D}_{linked}$
\end{algorithmic}
\end{algorithm}
Schema linking, the task of identifying the correct subset of tables and columns relevant to a user's question, is a cornerstone of any Text-to-SQL system~\cite{nl2sql-survey}. Existing methods~\cite{supersql, alpha-sql, chess} often treat this as a direct mapping task, which can be brittle when faced with complex schemas or ambiguous questions. An error at this stage is catastrophic, as an incomplete or incorrect schema makes generating a correct SQL query nearly impossible. This critical dependency is analogous to the role of a \textit{formal specification}~\cite{formal-specification} in software development; without an accurate specification, the final product is destined to fail. To address this, we introduce a \textit{Robust Schema Linking} module that, inspired by the principle of fault tolerance, combines multiple, diverse strategies to ensure the most accurate and complete schema is identified. Our overall process is detailed in Algorithm~\ref{alg:schema_linking}.

Our approach is a hybrid model that integrates three complementary linking techniques: (1)~\textbf{Direct Schema Linking}, (2)~\textbf{Reversed Schema Linking}, and (3)~\textbf{Value-based Schema Linking}.

\stitle{Direct Schema Linking.}
This method represents the most conventional approach, directly tasking the LLM with acting as a schema analysis agent. Given the user's question~$Q$, the full database schema~$\mathcal{D}$, and the retrieved relevant values~$\mathcal{M}_{retrieved}$, we prompt the LLM to explicitly list all relevant schema components. This process can be formalized as:
\begin{equation}
\mathcal{D}_{direct} = \text{LLM}_{\text{DirectLink}}(Q, \mathcal{D}, \mathcal{M}_{retrieved})
\end{equation}
While effective for unambiguous queries, its performance can degrade in complex scenarios, making it an unreliable standalone solution.

\stitle{Reversed Schema Linking.}
Inspired by the software engineering practice of \textit{reverse engineering}~\cite{reverse-engineering}, this technique reimagines the schema linking process. Instead of asking the LLM to first identify the schema, we prompt it to generate a draft SQL query directly, providing it with the full context including the relevant values. This approach is more effective because the task of generating SQL code aligns more closely with an LLM's pre-training on vast code corpora~\cite{askdata}. We then use a static parser to extract all schema components from the generated query~$S'$. Formally, the process is:
\begin{equation}
\mathcal{D}_{reversed} = \text{ParseSchema}(\text{LLM}_{\text{GenerateSQL}}(Q, \mathcal{D}, \mathcal{M}_{retrieved}))
\end{equation}
This ``answer-first'' approach allows the LLM to implicitly perform schema linking, often revealing components for complex joins or subqueries that a direct analysis might miss.

\stitle{Value-based Schema Linking.}
This technique provides an empirical, data-driven check to complement the model-driven approaches. It operates on the principle that if a column contains values highly similar to keywords in the question, that column is likely relevant. This method leverages the retrieved values~$\mathcal{M}_{retrieved}$ from the previous module. A column~$C_j$ is selected if any of its retrieved values has a similarity score with any question keyword~$k \in \mathcal{K}$ that exceeds a high-confidence threshold~$\theta_{val}$. This can be expressed as:
\begin{equation}
\begin{split}
\mathcal{D}_{value} = \{ C_j \in \mathcal{D} \mid & \exists v \in \mathcal{M}_{retrieved}[C_j], \exists k \in \mathcal{K} \\
& \text{ s.t. } \text{Sim}(v, k) > \theta_{val} \}
\end{split}
\end{equation}
This bottom-up method excels at resolving schema ambiguity. In cases where multiple columns are plausible candidates, it precisely identifies the correct one by grounding the selection in concrete data values instead of potentially misleading column names.

\stitle{Schema Union and Closure.}
The final step aggregates the results and ensures the relational integrity of the linked schema. First, we take the union of the schemas identified by all three methods:
\begin{equation}
\mathcal{D}_{union} = \mathcal{D}_{direct} \cup \mathcal{D}_{reversed} \cup \mathcal{D}_{value}
\end{equation}
\revA{However, since $\mathcal{D}_{union}$ is aggregated from independent strategies, it acts effectively as a collection of isolated schema nodes (tables). Without explicit edges (Foreign Keys) to connect them, the resulting schema graph is often disconnected, rendering valid JOIN operations impossible. To resolve this, we enforce relational closure as a mandatory connectivity safeguard.}
We parse all foreign key relationships in the database schema~$\mathcal{D}$. For any pair of tables~$(T_i, T_j)$~present in~$\mathcal{D}_{union}$, we automatically add the corresponding primary and foreign key columns that link them. 
\revA{This step transforms the set of isolated tables into a connected graph, providing a solid topological foundation for the subsequent SQL generation phase.}
This final, closed schema is defined as:
\begin{equation}
\mathcal{D}_{linked} = \text{EnforceClosure}(\mathcal{D}_{union}, \mathcal{D})
\end{equation}


\subsection{N-version Programming for SQL Generation}
\label{subsec:n_version}

\begin{algorithm}[t!]
\caption{N-version Programming for SQL Generation}
\label{alg:sql_generation}
\begin{algorithmic}[1]
\small
\Require User Question $Q$, Linked Schema $\mathcal{D}_{linked}$, Retrieved Values Map $\mathcal{M}_{retrieved}$
\Ensure  Set of Initial SQL Candidates $\mathcal{C}_{initial}$

\Statex \Comment{Step 1: Retrieve few-shot examples for the ICL generator}
\State $\mathcal{E}_{few-shot} \leftarrow \Func{RetrieveSimilarExamples}(Q)$

\Statex \Comment{Step 2: Execute the three SQL generators in parallel}
\State \textbf{begin parallel}
\State \quad $\mathcal{S}_{skel} \leftarrow \Func{LLM\_Skel}(Q, \mathcal{D}_{linked}, \mathcal{M}_{retrieved})$
\State \quad $\mathcal{S}_{icl} \leftarrow \Func{LLM\_ICL}(Q, \mathcal{D}_{linked}, \mathcal{M}_{retrieved}, \mathcal{E}_{few-shot})$
\State \quad $\mathcal{S}_{d\&c} \leftarrow \Func{LLM\_D\&C}(Q, \mathcal{D}_{linked}, \mathcal{M}_{retrieved})$
\State \textbf{end parallel}

\Statex \Comment{Step 3: Collect all generated SQLs into a candidate set}
\State $\mathcal{C}_{initial} \leftarrow \{\mathcal{S}_{skel}, \mathcal{S}_{icl}, \mathcal{S}_{d\&c}\}$ \Comment{and other samples if N > 1 per generator}

\State \textbf{return} $\mathcal{C}_{initial}$
\end{algorithmic}
\end{algorithm}
%


Upon establishing the query's specification in the \textit{Intent Scoping and Semantic Grounding} phase, the framework proceeds to the implementation stage: SQL generation. A single generation strategy, however, often struggles with the diversity of user queries~\cite{chase-sql}; a method effective for simple lookups may fail on complex analytical questions~\cite{din-sql}. To address this, \model instantiates a fault-tolerant strategy inspired by the software engineering principle of \textit{N-version programming}~\cite{n-version-programming}. Instead of relying on a single, monolithic generator, we deploy three distinct and independent SQL generators that run in parallel, each employing a different methodology. This engineered diversity significantly increases the likelihood of producing at least one correct candidate, enhancing the system's overall robustness. The entire workflow is detailed in Algorithm~\ref{alg:sql_generation}.

The inputs to this phase are the user question~$Q$, the linked schema~$\mathcal{D}_{linked}$, and the retrieved values~$\mathcal{M}_{retrieved}$. The three generators operate on this common set of inputs to produce a unified pool of initial SQL candidates~$\mathcal{C}_{initial}$.

\stitle{Skeleton-based SQL Generation.}
This generator is modeled after the \textit{top-down design} principle~\cite{top-down-design}, where a high-level plan is formulated before implementation details are filled in. This guides the LLM to think systematically, reducing structural errors. It involves three conceptual steps: component analysis, skeleton generation, and slot-filling. The entire process is encapsulated in a single call to the LLM, which is instructed to follow this reasoning path. We can formalize this as:
\begin{equation}
S_{skel} = \text{LLM}_{\text{Skel}}(Q, \mathcal{D}_{linked}, \mathcal{M}_{retrieved})
\end{equation}

\stitle{ICL-based SQL Generation.}
This generator leverages in-context learning (ICL), analogous to \textit{case-based reasoning}~\cite{case-based-reasoning}. By providing the LLM with relevant examples, we ground its generation in proven patterns.
\revA{To eliminate domain-specific bias, we adopt a masked retrieval strategy following DAIL-SQL~\cite{dail-sql}. Specifically, we mask table and column names in the user question with a special token (\ie \texttt{<MASK>}) to focus solely on structural logic. We then encode these masked questions using the \texttt{all-MiniLM-L6-v2}~\cite{reimers-2019-sentence-bert} model and retrieve the top-$K$ most similar examples from the training set of the corresponding benchmark (\eg BIRD-Train) to serve as few-shot demonstrations.}
This is formalized as:
\begin{equation}
S_{icl} = \text{LLM}_{\text{ICL}}(Q, \mathcal{D}_{linked}, \mathcal{M}_{retrieved}, \mathcal{E}_{few-shot})
\end{equation}
where~$\mathcal{E}_{few-shot}$ represents the set of retrieved few-shot examples.

\stitle{Divide-and-Conquer-based SQL Generation.}
For highly complex questions requiring nested logic, this generator implements the classic \textit{Divide and Conquer} paradigm. It breaks a large problem into smaller, manageable sub-problems that are solved recursively and then reassembled. This involves decomposing the question, solving each sub-question, and synthesizing the results into a single query. The process is formalized as:
\begin{equation}
S_{d\&c} = \text{LLM}_{\text{D\&C}}(Q, \mathcal{D}_{linked}, \mathcal{M}_{retrieved})
\end{equation}


\subsection{SQL Unit Testing and Revision via Tool-Chain}
\label{subsec:unit_testing}

\begin{algorithm}[t!]
\caption{SQL Unit Testing and Revision via Tool-Chain}
\label{alg:sql_revision}
\begin{algorithmic}[1]
\small
\Require A single SQL candidate $\mathcal{S}_{cand}$, Context ($Q, \mathcal{D}_{linked}, \mathcal{M}_{retrieved}$), Tool-Chain $\mathbb{C} = \{C_1, ..., C_N\}$
\Ensure  Revised SQL query $\mathcal{S}_{revised}$

\State $\mathcal{S}_{current} \leftarrow \mathcal{S}_{cand}$
\ForAll{checker $C_j \in \mathbb{C}$}
    \State $is\_valid, d_{err} \leftarrow \Func{C_j}(\mathcal{S}_{current})$
    \If{\textbf{not} $is\_valid$}
        \Statex \Comment{Error found, trigger revision and update the current SQL}
        \State $\mathcal{S}_{current} \leftarrow \Func{LLM}_{\text{Revise}}(Q, \mathcal{D}_{linked}, \mathcal{M}_{retrieved}, \mathcal{S}_{current}, d_{err})$
    \EndIf
\EndFor
\Statex \Comment{All checkers in the chain have been processed}
\State \textbf{return} $\mathcal{S}_{current}$
\end{algorithmic}
\end{algorithm}
%
\revA{LLM-generated SQL often contains critical errors (\eg incorrect \texttt{JOIN}s), yet LLMs struggle to self-correct due to inherent confirmation bias~\cite{nl2sql-bugs, share, self-correction-bias}. Analogous to the principle that code requires independent validation, we introduce \textit{SQL Unit Testing and Revision via Tool-Chain}. Drawing on \textit{Unit Testing}~\cite{unit-testing}, we treat functional SQL components (\eg \texttt{JOIN} logic, syntax) as testable ``units''. The \textit{Tool-Chain} externalizes verification through a deterministic chain of specialized checkers, functioning as an automated test suite to systematically detect errors and guide targeted revisions.}

\stitle{\revA{Tool-Chain of Checkers.}}

\begin{table}[t]
\small
\centering
\caption{\revA{Tool-chain of specialized checkers organized by verification stage.}}
\label{tab:checkers}
\begin{tabular}{p{0.18\columnwidth} p{0.28\columnwidth} p{0.44\columnwidth}}
\toprule
\textbf{Checker} & \textbf{Error Detection} & \textbf{Example Cases} \\
\midrule
\multicolumn{3}{c}{\textit{\textbf{Stage 1: Syntax \& Execution (Fail-Fast)}}} \\
\midrule
Syntax Checker
& Syntax validity and execution errors
& \texttt{SELECT col, COUNT(*)} (missing \texttt{GROUP BY}); \texttt{SELECT * FORM table} (typo) \\
\midrule
\multicolumn{3}{c}{\textit{\textbf{Stage 2: Logical Verification}}} \\
\midrule
JOIN Checker
& Non-standard or complex joins
& \texttt{ON T1.id = T2.id OR ...}; \texttt{ON col IN (SELECT ...)} \\
\cmidrule(lr){1-3}
ORDER-BY Checker
& Logical conflicts in sorting and limits
& \texttt{ORDER BY COUNT(*) LIMIT 1} (validate aggregation-based sorting) \\
\cmidrule(lr){1-3}
Time Checker
& Invalid temporal function usage
& \texttt{STRFTIME} vs. \texttt{DATETIME}; invalid date-format comparisons \\
\midrule
\multicolumn{3}{c}{\textit{\textbf{Stage 3: Quality \& Robustness}}} \\
\midrule
SELECT Checker
& Ambiguity removal
& \texttt{SELECT *} $\rightarrow$ \texttt{SELECT col\_a, col\_b} \\
\cmidrule(lr){1-3}
MaxMin Checker
& Efficiency optimization
& \texttt{WHERE col = (SELECT MAX(...))} $\rightarrow$ \texttt{ORDER BY col DESC LIMIT 1} \\
\cmidrule(lr){1-3}
NULL Checker
& Missing NULL guards
& Add \texttt{WHERE col IS NOT NULL} to sorting keys \\
\cmidrule(lr){1-3}
Result Checker
& Empty-result verification
& Queries returning zero rows (triggering re-evaluation of constraints) \\
\bottomrule
\end{tabular}
\end{table}

\revA{Our approach centers on a suite of specialized, deterministic programs we term ``Checker Tools''. These checkers are not applied randomly; they are orchestrated into a sequential \textit{``Syntax $\to$ Logic $\to$ Quality'' cascade}. This order is critical for a ``fail-fast'' verification: it ensures that fatal execution errors are resolved first before the system attempts to diagnose more subtle logical or stylistic discrepancies. The complete, ordered sequence is detailed in Table~\ref{tab:checkers}.}
%
\revA{The chain begins with the most fundamental validation: the \textit{Syntax Checker} ensures the query is syntactically valid and executable. \revA{Notably, this stage also acts as an implicit guard for aggregation errors (e.g., missing \texttt{GROUP BY}), as these typically trigger execution failures that are immediately captured and repaired.}
Once executable, the chain proceeds to \textit{Logical Verification}. The \textit{JOIN Checker} flags non-standard conditions, while the \textit{ORDER-BY Checker} validates sorting logic and corrects invalid patterns (e.g., misaligned \texttt{LIMIT} clauses).
Finally, the chain addresses \textit{Quality and Robustness} issues. The \textit{SELECT Checker} replaces ambiguous wildcards like \texttt{SELECT *} with specific column names. Concurrently, the \textit{NULL Checker} and \textit{Result Checker} inject guards for potential \texttt{NULL} issues and flag queries that produce empty results, ensuring the final output is not only correct but also practically useful.}

\stitle{The Sequential Check-and-Revise Process.}
Our framework implements an efficient, single-pass ``check-and-revise'' process, detailed in Algorithm~\ref{alg:sql_revision}. Each initial SQL candidate, $\mathcal{S} \in \mathcal{C}_{initial}$, is passed through the \textit{Tool-Chain of Checkers} exactly once. The process for a single candidate $\mathcal{S}_{cand}$ is as follows:
\begin{enumerate}[leftmargin=*, topsep=0pt, itemsep=0pt, parsep=0pt]
    \item The candidate $\mathcal{S}_{cand}$ is sequentially evaluated by each checker in the tool-chain, starting with the first.
    \item If a checker detects an error, the process is momentarily paused. The checker generates a specific error report and an actionable \textit{correction directive}, $d_{err}$. For example, if the \texttt{NULL Checker} finds that a column in an \texttt{ORDER BY} clause could contain \texttt{NULL} values, the directive would be a clear instruction such as: {``The ordering column \texttt{[column\_name]} may contain NULLs. Add a \texttt{WHERE {[column\_name]} IS NOT NULL} condition to ensure correct sorting.''}

    \item The LLM is then invoked in a special ``revision mode''. It receives the original context, the faulty SQL $\mathcal{S}_{cand}$, and the explicit directive $d_{err}$ from the checker. The LLM's task is not to find the error, but to fix it based on the directive. It is formalized as:
        \begin{equation}
        \mathcal{S}_{revised} = \text{LLM}_{\text{Revise}}(Q, \mathcal{D}_{linked}, \mathcal{M}_{retrieved}, \mathcal{S}_{cand}, d_{err})
        \end{equation}
    \item The newly revised query, $\mathcal{S}_{revised}$, replaces $\mathcal{S}_{cand}$, and the evaluation continues with the next checker in the chain.
    \item The process terminates once the query has been evaluated by all checkers in the chain.
\end{enumerate}
This external, tool-guided debugging process is significantly more reliable than unconstrained LLM self-correction. The final output of this phase is a set of revised SQL candidates, $\mathcal{C}_{revised}$, which have been rigorously vetted and have a substantially higher probability of being correct.

\subsection{Confidence-aware SQL Selection}
\label{subsec:sql_selection}

\begin{algorithm}[t!]
\caption{Confidence-aware SQL Selection}
\label{alg:sql_selection}
\begin{algorithmic}[1]
\small
\Require Revised SQL Candidates $\mathcal{C}_{revised}$, Context ($Q, \mathcal{D}_{linked}, \mathcal{M}_{retrieved}$), Threshold $\theta_{conf}$
\Ensure  The Final SQL Query $\mathcal{S}_{final}$

\Statex \Comment{Step 1: Execute all candidates and cluster the results}
\State $\mathcal{R} \leftarrow \Func{ExecuteAll}(\mathcal{C}_{revised})$
\State $\{\text{Cluster}_1, ..., \text{Cluster}_M\}, \{\mathcal{S}_1, ..., \mathcal{S}_M\} \leftarrow \Func{ClusterAndRank}(\mathcal{R})$

\Statex \Comment{Step 2: Calculate confidence for the top-ranked candidate}
\State $Conf(\mathcal{S}_1) \leftarrow \frac{|\text{Cluster}_1|}{|\mathcal{C}_{revised}|}$

\Statex \Comment{Step 3: Confidence-Gated Selection Path}
\If{$Conf(\mathcal{S}_1) > \theta_{conf}$}
    \State $\mathcal{S}_{final} \leftarrow \mathcal{S}_1$ \Comment{High-Confidence Shortcut}
\Else
    \State Let $\{\mathcal{S}_1, ..., \mathcal{S}_K\}$ be the top-K candidates \Comment{Low-Confidence Full Review}
    \ForAll{candidate $\mathcal{S}_i$ in $\{\mathcal{S}_1, ..., \mathcal{S}_K\}$}
        \State $Conf(\mathcal{S}_i) \leftarrow \frac{|\text{Cluster}_i|}{|\mathcal{C}_{revised}|}$
        \State $WinRate(\mathcal{S}_i) \leftarrow \Func{LLM-PairwiseVoting}(\{\mathcal{S}_1, ..., \mathcal{S}_K\})$ \Comment{Using Eq.~\ref{eq:winrate}}
        \State $Score(\mathcal{S}_i) \leftarrow Conf(\mathcal{S}_i) \times WinRate(\mathcal{S}_i)$
    \EndFor
    \State $\mathcal{S}_{final} \leftarrow \arg\max_{\mathcal{S}_i} Score(\mathcal{S}_i)$
\EndIf

\State \textbf{return} $\mathcal{S}_{final}$
\end{algorithmic}
\end{algorithm}
%
The preceding phases of our framework produce a set of high-quality, revised SQL candidates, $\mathcal{C}_{revised}$. However, these candidates may not be identical and could yield different execution results. The most common approach for selecting a final query is \textit{self-consistency}~\cite{opensearch-sql, alpha-sql, dail-sql}, where all candidates are executed, and the query corresponding to the most frequent result is chosen. This method, while a strong baseline, has a critical flaw: \textit{the most popular answer is not always the correct one, especially for complex problems where multiple generation paths might converge on the same plausible but incorrect logic}.
This final challenge mirrors the release stage in a software lifecycle, which is governed by a \textit{Quality Gate}~\cite{quality-gate}. A quality gate's purpose is not merely to accept the most-voted-for version, but to enforce a set of objective quality criteria before a product is released. Similarly, our \textit{Confidence-aware SQL Selection} phase acts as this quality gate for the generated SQL.
It overcomes the flaws of simple majority voting by using the initial vote's confidence as a quality metric to guide a more reliable, adaptive selection process.

\stitle{Execution-based Clustering and Confidence Estimation.}
The process begins by executing every revised SQL candidate $\mathcal{S} \in \mathcal{C}_{revised}$ on the database. The resulting datasets are then clustered, such that queries producing identical results are grouped together. These clusters are ranked based on their size (\ie the number of SQL queries they contain). For each of the top-$K$ candidates $\{\mathcal{S}_1, \mathcal{S}_2, ..., \mathcal{S}_K\}$, representing the top-$K$ largest clusters, we calculate an execution-based confidence score. The confidence of a candidate $\mathcal{S}_i$ is the proportion of total queries that belong to its cluster:
\begin{equation}
Conf(\mathcal{S}_i) = \frac{|\text{Cluster}_i|}{|\mathcal{C}_{revised}|}
\end{equation}
This score, particularly $Conf(\mathcal{S}_1)$, serves as a strong indicator of the query's likely correctness, as shown by the high correlation in Figure~\ref{fig:ex-vs-confidence}.

\stitle{Confidence-Gated Selection.}
Based on the confidence of the top-ranked candidate $\mathcal{S}_1$, our framework follows one of two distinct paths, as detailed in Algorithm~\ref{alg:sql_selection}.

\etitle{High-Confidence Shortcut.}
If the confidence score $Conf(\mathcal{S}_1)$ exceeds a predefined high-confidence threshold $\theta_{conf}$, we conclude that there is overwhelming agreement among the generated candidates. In this scenario, we directly select $\mathcal{S}_1$ as the final query. This shortcut avoids unnecessary and costly LLM invocations for cases where the answer is already clear, providing a practical trade-off between accuracy and efficiency.

\etitle{Low-Confidence Full Review.}
If $Conf(\mathcal{S}_1) < \theta_{conf}$, it signifies substantial ambiguity or disagreement among the candidates, making the top choice unreliable. It then triggers a full review pipeline.

\stab\revA{(1) \textbf{Execution-Guided Prompting.}}
\revA{Instead of a neutral comparison, we construct an asymmetric prompt guided by execution consensus. We identify the top-$K$ execution clusters and randomly select one representative SQL from each cluster to form the candidate set.
The prompt explicitly injects the execution confidence as context (e.g., informing the LLM that Candidate A has higher execution confidence than Candidate B). 
We instruct the LLM to treat this high-confidence consensus as a prior, trusting it unless specific \textit{logical evidence}---such as a contradiction between the generated SQL and the user's natural language requirement---demonstrably proves it factually incorrect.}

\stab\revA{(2) \textbf{Perform Pairwise Adjudication.}}
\revA{Next, the LLM performs pairwise comparisons for the top-$K$ candidates. To ensure reliability and mitigate the inherent stochasticity of LLM outputs, we employ a \textit{self-consistency} mechanism. For each pair $(\mathcal{S}_i, \mathcal{S}_j)$, we sample three independent judgments and define the final stable vote, $V(\mathcal{S}_i, \mathcal{S}_j)$, as the majority outcome. The vote $V(\mathcal{S}_i, \mathcal{S}_j)$ yields 1 if $\mathcal{S}_i$ is superior and 0 if $\mathcal{S}_j$ is superior. From these consistent pairwise results, we compute an aggregate win rate for each candidate $\mathcal{S}_i$ as its average score against all other competitors in the top-$K$ set:}
    \begin{equation}
    \label{eq:winrate}
    WinRate(\mathcal{S}_i) = \frac{1}{K-1} \sum_{j=1, j \neq i}^{K} V(\mathcal{S}_i, \mathcal{S}_j)
    \end{equation}
    
\stab(3)  \textbf{Calculate Final Score.} Finally, the decision is based on a confidence-aware score that combines the prior execution-based confidence with the LLM-adjudicated win rate:
    \begin{equation}
    Score(\mathcal{S}_i) = Conf(\mathcal{S}_i) \times WinRate(\mathcal{S}_i)
    \end{equation}
The query with the highest overall score, $\arg\max_{\mathcal{S}_i} Score(\mathcal{S}_i)$, is selected as the final SQL, $\mathcal{S}_{final}$.

\subsection{Workflow Optimization}
\label{subsec:optimization}
While \model's multi-stage architecture ensures robustness, we incorporate three key optimizations to maintain efficiency and computational costs, preventing prohibitive latency or expense.

\stitle{Efficient Prompting Strategy.}
Instead of sequential, costly LLM API calls for multi-step reasoning (\eg planning), we use a single, sophisticated prompt for each generator. This prompt includes a ``chain-of-thought'' instruction, directing the LLM to perform the entire logical sequence internally and output only the final SQL in one call. This preserves structured reasoning benefits while eliminating the associated latency and token overhead.

\stitle{Parallel Execution.}
To reduce end-to-end latency, our framework heavily parallelizes independent tasks. Critical components are executed concurrently, including: multi-column value retrieval, all three schema linking strategies, the N-version SQL generators, and the parallel revision of each SQL candidate.

\stitle{Conditional Execution.}
To minimize unnecessary LLM invocations, \model employs conditional execution at two critical stages. First, during \textit{SQL Unit Testing and Revision}, the LLM is only called for revision if a checker tool detects an error. Second (Section~\ref{subsec:sql_selection}), LLM-based pairwise adjudication is only triggered in low-confidence scenarios.

\section{Experiments}
\label{sec:exp}

\revA{This section evaluates the performance and reliability of \model. We first detail the experimental setup, followed by a comparative analysis against state-of-the-art methods. Finally, we conduct ablation studies and efficiency analyses to validate the contribution of individual components.}

\subsection{Experimental Setup}
\label{subsec:exp_setup}

\stitle{Datasets.}
\revB{We evaluate the performance of \model on three challenging Text-to-SQL benchmarks: \textbf{BIRD}~\cite{bird}, \textbf{Spider}~\cite{spider}, and the newly released \textbf{Spider 2.0}~\cite{spider2}}.
BIRD is a large-scale benchmark designed to mirror complex, real-world scenarios. It contains 12,751 unique question-SQL pairs across 95 databases from over 37 professional domains. Its databases are notably large and feature messy data and intricate schemas, making it a difficult test for grounding and robustness.
Spider is a foundational large-scale, cross-domain benchmark in the field. It consists of 10,181 questions and 5,693 unique, complex SQL queries across 200 databases covering 138 different domains.
\revB{\textbf{Spider 2.0} represents a significant leap towards enterprise-grade challenges. Unlike its predecessor, it introduces complex schema and diverse dialects (e.g., BigQuery, Snowflake) to simulate real-world data warehousing environments. It is divided into two subsets: \textit{Lite} and \textit{Snow}.
Following prior works~\cite{chase-sql, xiyan-sql, rsl-sql, reforce}, we use the development set of BIRD (BIRD-Dev), the test set of Spider (Spider-Test), and both the \textit{Snow} and \textit{Lite} subsets of Spider 2.0 for our main evaluation.}


\stitle{Evaluation Metrics.}
Following prior works~\cite{chase-sql, xiyan-sql, rsl-sql}, our primary evaluation metric is \textbf{Execution Accuracy (EX)}. 
\revB{For BIRD and Spider, a generated SQL is considered correct if its execution result is strictly equivalent to the ground truth.
For Spider 2.0, we adhere to its official evaluation protocol~\cite{spider2}, which employs a relaxed comparison standard that permits redundant columns in the result set, acknowledging the practical needs of data warehousing environments.}
To measure the potential of our N-version Programming for SQL Generation module, we report the \textbf{Upper-bound EX}, which is the execution accuracy an oracle would achieve by always selecting the correct SQL from the generated candidates~\cite{chase-sql}. For evaluating the Robust Schema Linking module, we use \textbf{Table/Column Recall}, the proportion of ground-truth tables and columns correctly identified. Finally, to assess practical efficiency, we measure \textbf{Token Cost}, corresponding to the number of tokens processed by LLMs.

\stitle{Baselines.}
We compare \model against SOTA baselines from two paradigms (Table~\ref{tab:performance}). 
The first is \textbf{fine-tuning-based methods}, which are trained on in-domain data, including strong competitors like XiYan-SQL~\cite{xiyan-sql}, CHASE-SQL~\cite{chase-sql}, and OmniSQL~\cite{omnisql}. 
The second is \textbf{prompting-based methods}, which, like \model, are training-free. This group includes methods like Alpha-SQL~\cite{alpha-sql}, RSL-SQL~\cite{rsl-sql}, and CHESS~\cite{chess}.
\revB{Furthermore, for the Spider 2.0 evaluation, we compare against specialized agentic frameworks designed for enterprise-grade complexity, including Spider-Agent~\cite{spider2}, ReFoRCE~\cite{reforce} and LinkAlign~\cite{linkalign}.}

\stitle{Implementation Details.}
All experiments are conducted on an Ubuntu 22.04.3 LTS server with 512GB of RAM and dual 40-core Intel(R) Xeon(R) Platinum 8383C CPUs. We deploy all open-source LLMs locally using the vllm~\cite{vllm} framework on 8 NVIDIA A100 GPUs, each with 80GB of memory, and accelerate inference using a tensor parallelism of 8. 
To validate the robustness and generalizability of our framework, we integrated \model with three distinct models from two leading series: \texttt{Gemma3-27B-Instruct}~\cite{gemma3}, \texttt{Qwen2.5-Coder-32B-Instruct}~\cite{qwen2.5-coder}, and \texttt{Qwen3-Coder-30B-A3B-Instruct}~\cite{qwen3}. 
\revB{Additionally, for the Spider 2.0 benchmark, which requires extreme reasoning capabilities to navigate complex enterprise schemas, we adopted the \texttt{DeepSeek-R1} model~\cite{deepseek-r1}, consistent with the configuration of comparable baselines.}
Unless otherwise specified, the pipeline is configured as follows. For Semantic Value Retrieval, where we retrieve up to the top-5 most similar values for each TEXT-type column, we use the \texttt{Qwen3-Embedding-0.6B}~\cite{qwen3-embedding} model, and the similarity threshold ($\theta_{val}$) for Value-based Schema Linking is set to 0.98. The draft SQL for Reversed Schema Linking is produced by the ICL-based generator. For each LLM sub-task (\eg Direct Schema Linking), we set the sampling budget to 8 with a sampling temperature of 0.7. Finally, in the Confidence-aware SQL Selection phase, which adjudicates between the top-2 ranked queries in low-confidence scenarios, the confidence shortcut threshold ($\theta_{conf}$) is set to 0.6.

\subsection{Overall Performance}
\label{subsec:overall_perf}

\begin{table*}[t!]
\centering
\caption{Performance comparison on BIRD-Dev and Spider-Test datasets.}
\label{tab:performance}
\small
\resizebox{\textwidth}{!}{
\begin{tabular}{@{}llccc@{}}
\toprule
\multicolumn{1}{l|}{\textbf{Methods}} &
  \textbf{Model} &
  \multicolumn{1}{c|}{\textbf{\# Parameters}} &
  \textbf{BIRD-EX (\%)} &
  \textbf{Spider-EX (\%)} \\ \midrule
\multicolumn{5}{c}{\textit{\textbf{Fine-tuning-based Baselines}}}                                                             \\ \midrule
\multicolumn{1}{l|}{SENSE~\cite{sense}}               & CodeLLaMA-13B              & \multicolumn{1}{c|}{$\sim$13B}          & 55.5 & 86.6 \\
\multicolumn{1}{l|}{SFT CodeS~\cite{codes}}           & CodeS-15B                  & \multicolumn{1}{c|}{$\sim$15B}          & 58.5 & -    \\
\multicolumn{1}{l|}{Distillery~\cite{distillery}}          & GPT-4o                     & \multicolumn{1}{c|}{\textgreater{}200B} & 67.2 & -    \\
\multicolumn{1}{l|}{XiYanSQL-QwenCoder~\cite{xiyan-sql}} &
  Qwen2.5-Coder-32B-Instruct &
  \multicolumn{1}{c|}{$\sim$32B} &
  67.1 &
  88.4 \\
\multicolumn{1}{l|}{BASE-SQL~\cite{base-sql}}            & Qwen2.5-Coder-32B-Instruct & \multicolumn{1}{c|}{$\sim$32B}          & 67.5 & 88.9 \\
\multicolumn{1}{l|}{OmniSQL~\cite{omnisql}}             & Qwen2.5-Coder-32B-Instruct & \multicolumn{1}{c|}{$\sim$32B}          & 67.0 & 89.8 \\
\multicolumn{1}{l|}{CSC-SQL~\cite{csc-sql}}             & XiYanSQL-QwenCoder-32B     & \multicolumn{1}{c|}{$\sim$32B}          & 71.3 & -    \\
\multicolumn{1}{l|}{CHASE-SQL~\cite{chase-sql}} &
  Gemini-1.5-Pro + Gemini-1.5-Flash &
  \multicolumn{1}{c|}{\textgreater{}200B} &
  73.0 &
  87.6 \\
\multicolumn{1}{l|}{XiYan-SQL~\cite{xiyan-sql}} &
  GPT-4o + Qwen2.5-Coder-32B-Instruct &
  \multicolumn{1}{c|}{\textgreater{}200B} &
  73.3 &
  89.7 \\ \midrule
\multicolumn{5}{c}{\textit{\textbf{Prompting-based Baselines}}}                                                               \\ \midrule
\multicolumn{1}{l|}{DIN-SQL~\cite{din-sql}}             & GPT-4                      & \multicolumn{1}{c|}{\textgreater{}175B} & 50.7 & 85.3 \\
\multicolumn{1}{l|}{DAIL-SQL~\cite{dail-sql}}            & GPT-4                      & \multicolumn{1}{c|}{\textgreater{}175B} & 55.9 & 86.6 \\
\multicolumn{1}{l|}{SuperSQL~\cite{supersql}}            & GPT-4                      & \multicolumn{1}{c|}{\textgreater{}175B} & 58.5 & -    \\
\multicolumn{1}{l|}{RSL-SQL~\cite{rsl-sql}}             & GPT-4o                     & \multicolumn{1}{c|}{\textgreater{}200B} & 67.2 & 87.9 \\
\multicolumn{1}{l|}{CHESS (IR,CG,UT)~\cite{chess}} &
  Gemini-1.5-Pro &
  \multicolumn{1}{c|}{\textgreater{}200B} &
  68.3 &
  - \\
\multicolumn{1}{l|}{OpenSearch-SQL (v2)~\cite{opensearch-sql}} & GPT-4o                     & \multicolumn{1}{c|}{\textgreater{}200B} & 69.3 & 87.1 \\
\multicolumn{1}{l|}{Alpha-SQL~\cite{alpha-sql}}           & Qwen2.5-Coder-32B-Instruct & \multicolumn{1}{c|}{$\sim$32B}          & 69.7 & -    \\ \midrule
\multicolumn{5}{c}{\textit{\textbf{Our \model (Prompting-based)}}}                                                         \\ \midrule
\multicolumn{1}{l|}{\textbf{\model}} &
  \textbf{Gemma3-27B-Instruct} &
  \multicolumn{1}{c|}{$\sim$27B} &
  \textbf{71.1} &
  \textbf{88.9} \\
\multicolumn{1}{l|}{\textbf{\model}} &
  \textbf{Qwen2.5-Coder-32B-Instruct} &
  \multicolumn{1}{c|}{$\sim$32B} &
  \textbf{70.9} &
  \textbf{88.7} \\
\multicolumn{1}{l|}{\textbf{\model}} &
  \textbf{Qwen3-Coder-30B-A3B-Instruct} &
  \multicolumn{1}{c|}{$\sim$30B} &
  \textbf{73.5} &
  \textbf{89.8} \\ \bottomrule
\end{tabular}%
}
\end{table*}

\begin{table}[t!]
\centering
\caption{\revB{Execution Accuracy (\%) on Spider 2.0 Benchmark. All methods utilize DeepSeek-R1 as the backend LLM to ensure a fair comparison.}}
\label{tab:spider2}
\small
\begin{tabular}{l|cc}
\toprule
\textbf{Methods} & \textbf{Spider 2.0-Snow} & \textbf{Spider 2.0-Lite} \\
\midrule
DIN-SQL~\cite{din-sql} & 0.0 & 4.8 \\
DAIL-SQL~\cite{dail-sql} & 6.6 & 8.8 \\
Spider-Agent~\cite{spider2} & 10.6 & 13.7 \\
ReFoRCE~\cite{reforce} & 38.0 & 29.6 \\
RSL-SQL~\cite{rsl-sql} & - & 30.5 \\
LinkAlign~\cite{linkalign} & - & 33.1 \\
\midrule
\textbf{\model (Ours)} & \textbf{50.5} & \textbf{38.2} \\
\bottomrule
\end{tabular}
\end{table}

\stitle{RQ1: How does \model perform against existing state-of-the-art methods on challenging Text-to-SQL benchmarks?}

\revB{To answer this, we present the main performance of \model on the BIRD-Dev, Spider-Test, and Spider 2.0 datasets in Table~\ref{tab:performance} and Table~\ref{tab:spider2}, comparing it against a wide range of state-of-the-art fine-tuning, prompting-based, \revB{and agentic} methods. The results clearly demonstrate that our software-grounded framework establishes a new benchmark for Text-to-SQL generation.}
\revB{Specifically, when integrated with the \texttt{Qwen3-Coder-30B-A3B} model, \model achieves an execution accuracy of \textbf{73.5\%} on BIRD-Dev. This result not only significantly outperforms all existing prompting-based baselines but, more remarkably, it also surpasses the leading fine-tuning-based systems like XiYan-SQL (73.3\%) and CHASE-SQL (73.0\%). It is crucial to note that these competing methods rely on substantially larger and often proprietary models (\eg GPT-4o and Gemini-1.5-Pro).}
%
\revB{On the official BIRD held-out Test set, \model achieves \textbf{75.07\%} execution accuracy with open-source models. Furthermore, scaling up to \texttt{Gemini-3-Pro} yields \textbf{76.58\%} (Global Rank 5th) even with minimal sampling ($N=1$).}
\revB{On the Spider-Test dataset, \model with \texttt{Qwen3-Coder-30B-A3B} achieves an execution accuracy of \textbf{89.8\%}. This performance matches the best fine-tuned method, OmniSQL, and surpasses other strong competitors like XiYan-SQL (89.7\%). This demonstrates that our prompting-based framework can attain the same level of accuracy as highly specialized, fine-tuned models on this foundational benchmark without incurring any training costs.
}
\revB{Moreover, on the newly released enterprise-grade Spider 2.0 benchmark (Table~\ref{tab:spider2}), \model demonstrates superior robustness in handling complex, real-world constraints. On the rigorous ``Snow'' subset, \model achieves \textbf{50.5\%} execution accuracy, surpassing the specialized agentic framework ReFoRCE (38.0\%) by a large margin (+12.5\%). Notably, traditional prompting methods like DIN-SQL fail completely on this task (0.0\%), highlighting that our structured engineering lifecycle is essential for navigating the ambiguity of realistic data warehousing scenarios.}
\revB{Furthermore, the consistently high performance of \model across all three tested open-source models---71.1\% with \texttt{Gemma3-27B}, 70.9\% with \texttt{Qwen2.5-Coder-32B}, and 73.5\% with \texttt{Qwen3-Coder-30B} on BIRD, which underscores the robustness and generalizability of our framework (Table~\ref{tab:fair-comparison}). }

\stitle{RQ2: Is the performance gain of \model attributable to its architectural design rather than the power of the underlying base model?}

\begin{table}[t]
\small
\centering
\caption{Performance comparison using the same LLM backbone. 
Gemma3-27B, Qwen2.5-32B, and Qwen3-30B denote 
Gemma3-27B-Instruction, Qwen2.5-Coder-32B-Instruct, and 
Qwen3-Coder-30B-A3B-Instruct, respectively.}
\label{tab:fair-comparison}
\setlength{\tabcolsep}{4pt}
\renewcommand{\arraystretch}{1.05}
\begin{tabular}{l rr rr rr}
\toprule
\textbf{Methods}
& \multicolumn{2}{c}{\textbf{Gemma3-27B}}
& \multicolumn{2}{c}{\textbf{Qwen2.5-32B}}
& \multicolumn{2}{c}{\textbf{Qwen3-30B}} \\
\cmidrule(lr){2-3}\cmidrule(lr){4-5}\cmidrule(lr){6-7}
& \textbf{EX} & \textbf{$\Delta$}
& \textbf{EX} & \textbf{$\Delta$}
& \textbf{EX} & \textbf{$\Delta$} \\
\midrule
CoT-Baseline            & 54.2 & --    & 61.3 & --    & 61.7 & --    \\
CHESS (IR, CG, UT)      & 66.1 & +11.9 & 67.7 & +6.4  & 67.9 & +6.2  \\
Alpha-SQL               & 68.6 & +14.4 & 69.7 & +8.4  & 71.2 & +9.5  \\
\textbf{\model\ (Ours)} & \textbf{71.1} & \textbf{+16.9}
                         & \textbf{70.9} & \textbf{+9.6}
                         & \textbf{73.5} & \textbf{+11.8} \\
\bottomrule
\end{tabular}
\end{table}

To isolate our framework's contribution from the base model's intrinsic capabilities, we evaluated it against leading prompting-based frameworks on identical open-source models (Table~\ref{tab:fair-comparison}). The results confirm that \model's architecture consistently provides the most substantial performance improvement. For instance, on \texttt{Gemma3-27B-Instruct}, \model improves the CoT baseline by +16.9\%, significantly outpacing the gains from CHESS (+11.9\%) and Alpha-SQL (+14.4\%). This trend holds across all tested models, culminating in a state-of-the-art 73.5\% EX on \texttt{Qwen3-Coder-30B-A3B} (+11.8\% gain).

\subsection{Robust Schema Linking Analysis}

\begin{table}[t]
\small
\centering
\caption{Schema linking analysis on the BIRD-Dev dataset.}
\label{tab:schema-linking}
\setlength{\tabcolsep}{4pt}
\renewcommand{\arraystretch}{1.05}
\begin{tabular}{lccc}
\toprule
\multirow{2}{*}{\textbf{Schema Linking Methods}} 
& \textbf{Table} 
& \textbf{Column} 
& \textbf{Avg.} \\
& \textbf{Recall (\%)} 
& \textbf{Recall (\%)} 
& \textbf{Tokens} \\
\midrule
No Schema Linking              & --           & --           & 5486.2 \\
Direct Schema Linking          & 94.2         & 80.9         & 454.5  \\
Reversed Schema Linking        & 97.0         & 94.0         & 495.9  \\
Value Schema Linking           & 47.3         & 18.0         & 262.0  \\
\midrule
\textbf{Robust Schema Linking} & \textbf{98.1} & \textbf{95.4} & \textbf{627.4} \\
\bottomrule
\end{tabular}
\end{table}

\stitle{RQ3: How do the individual components of our Robust Schema Linking module contribute to its overall effectiveness and efficiency?}

\begin{figure}[t!]
\begin{tcolorbox}[
  arc=2mm,
  colback=black!5!white,
  colframe=black!75!white,
  boxrule=1pt,
  boxsep=0mm,
]
\small
\textbf{Question:} Which accounts placed orders for \textit{``household payment''} in Pisek?

\textbf{Evidence:} \texttt{k\_symbol = `SIPO'} refers to household payment.

\tcbline

\textbf{LLM-based Linking (Direct \& Reversed):}
\begin{itemize}
    \item[\small\texttt >] \textit{Found:} \texttt{account.account\_id}, \texttt{district.district\_id}, \texttt{order.k\_symbol}, ...
    \item[\textcolor{red}{\sffamily\bfseries X}] \textbf{Missed:} The essential column \texttt{trans.k\_symbol}.
\end{itemize}

\textbf{Value-based Linking:}
\begin{itemize}[nosep]
    \item[\textcolor{green!50!black}{\sffamily\bfseries \checkmark}] \textbf{Recovered:} \texttt{trans.k\_symbol} by linking to value \texttt{`SIPO'}.
\end{itemize}

\begin{lstlisting}[style=sqlstyle]
-- Gold SQL
SELECT DISTINCT T2.account_id FROM trans AS T1 JOIN account AS T2 ON T1.account_id = T2.account_id JOIN district AS T3 ON T2.district_id = T3.district_id WHERE T1.k_symbol = 'SIPO' AND T3.A2 = 'Pisek'
\end{lstlisting}
\end{tcolorbox}
\caption{A case study from BIRD-Dev (QID: 142) where Value-based Linking recovers a critical column (\texttt{trans.k\_symbol}) missed by purely LLM-based linking methods.}
\label{fig:case_study_value_linking}
\end{figure}

We analyzed our \textit{Robust Schema Linking} module's components on BIRD-Dev (Table~\ref{tab:schema-linking}), measuring schema recall and token efficiency. The full approach achieves the highest recall (\textbf{98.1\%} table, \textbf{95.4\%} column), providing a critical foundation for SQL generation. It also dramatically boosts efficiency, reducing the input context by 9$\times$ from 5486.2 to 627.4 tokens compared to using the full schema.

Analyzing the components reveals complementary strengths. Our novel \textit{Reversed Schema Linking} (97.0\% table, 94.0\% column) significantly outperforms \textit{Direct Schema Linking}, especially in column recall (+13.1\%). This validates our hypothesis that prompting for a draft query is a more natural and effective reasoning method for LLMs than explicit extraction.

Finally, \textit{Value-based Schema Linking}, while low in overall recall, is not a standalone linker. It acts as a precise mechanism to link columns via their data values, catching omissions from other methods. For instance (Figure~\ref{fig:case_study_value_linking}), it correctly identified \texttt{trans.k\_symbol} by linking the query's ``household payment'' to the data value \texttt{`SIPO'}. This bridged a semantic gap that schema-level reasoners (Direct and Reversed) missed, as they were confused by an ambiguous \texttt{k\_symbol} name. This recovery proves the value-based method's essential role in our fault-tolerant design.

\subsection{Analysis on N-Version Programming for SQL Generation}

\label{subsec:analysis_n_version}

\begin{table}[t]
\small
\centering
\caption{Analysis of SQL generation methods on the BIRD-Dev dataset.}
\label{tab:sql-generation}
\setlength{\tabcolsep}{4pt}
\renewcommand{\arraystretch}{1.05}
\begin{tabular}{lcc}
\toprule
\textbf{SQL Generation Method} & \textbf{EX (\%)} & \textbf{UB-EX (\%)} \\
\midrule
Skeleton-based SQL Generation & 69.2 & 77.9 \\
ICL-based SQL Generation      & 70.9 & 78.3 \\
D\&C-based SQL Generation     & 70.3 & 78.5 \\
\midrule
\textbf{N-version Programming for SQL Generation} & \textbf{71.7} & \textbf{81.1} \\
\bottomrule
\end{tabular}
\end{table}

\begin{figure}[t!]
\centering
\begin{subfigure}{0.48\columnwidth}
    \centering
    \includegraphics[width=\textwidth]{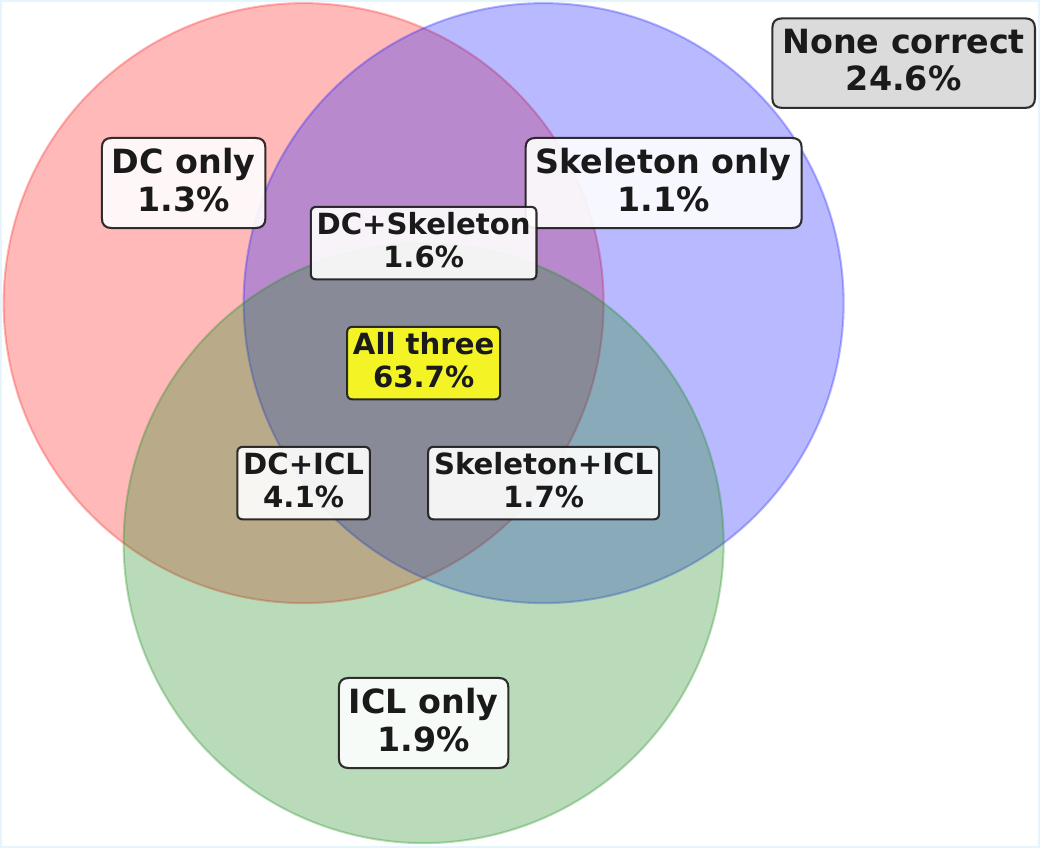}
    \caption{EX Correctness Overlap.}
    \label{fig:sc_ex_venn}
\end{subfigure}
\hfill
\begin{subfigure}{0.48\columnwidth}
    \centering
    \includegraphics[width=\textwidth]{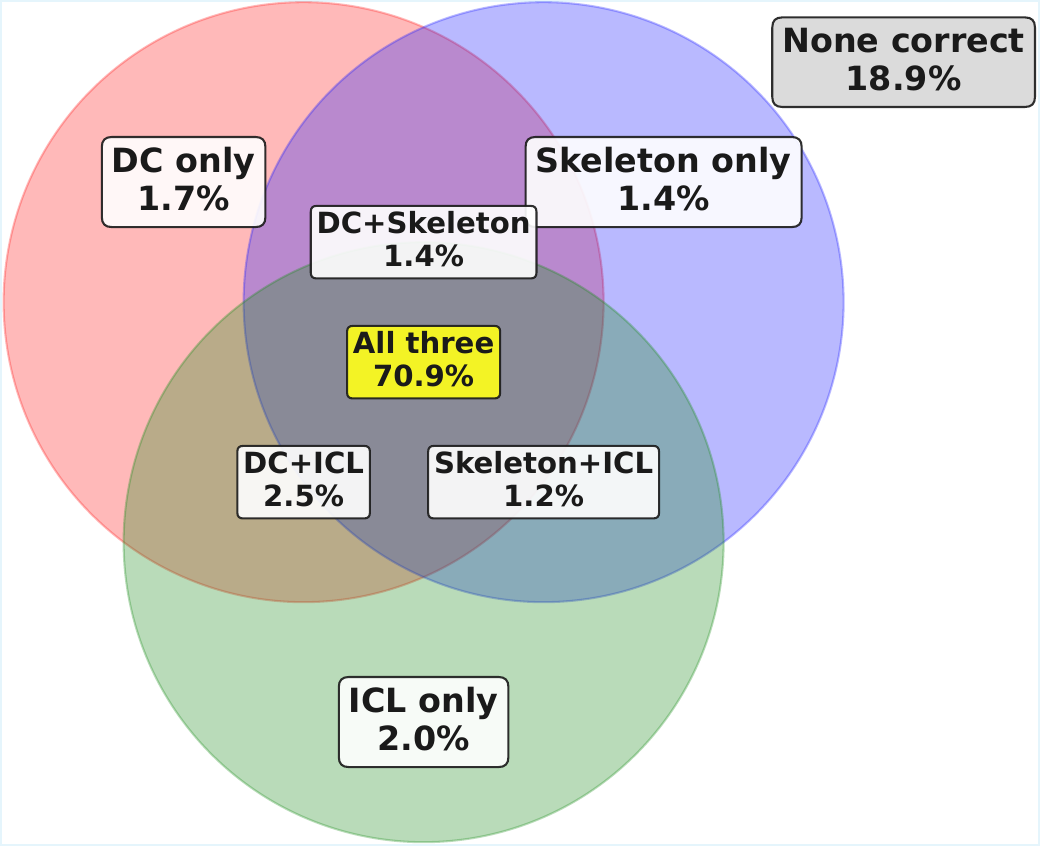}
    \caption{UB-EX Correctness Overlap.}
    \label{fig:ub_venn}
\end{subfigure}
\caption{Correctness overlap analysis of three SQL generation methods on BIRD-Dev dataset.}
\label{fig:venn_diagrams}
\end{figure}

\stitle{RQ4: How do the different SQL generators contribute to the overall performance, and does the N-version programming approach provide a tangible benefit?}

We analyzed our three SQL generators' individual and combined performance (EX and UB-EX) on BIRD-Dev (Table~\ref{tab:sql-generation}, Figure~\ref{fig:venn_diagrams}). While all generators perform well individually (peaking at 70.9\% EX), combining them in our N-version module already yields a 71.7\% EX. The true strength of this approach, however, lies in its potential: the combined UB-EX reaches \textbf{81.1\%}, a significant +2.6\% gain over the best individual generator's potential (78.5\%).
This UB-EX increase confirms that the generators possess crucial diversity, solving different subsets of problems where one's failure is covered by another's success. The overlap analysis (Figure~\ref{fig:ub_venn}) corroborates this: while 70.9\% of correct answers are found by all three generators, a significant \textbf{10.2\%} are solved by only one or two. This proves the generators are not redundant. This engineered diversity—finding a correct answer for 81.1\% of questions—validates our fault-tolerant design and provides the rich candidate pool essential for achieving state-of-the-art performance

\stitle{\revA{RQ5: Is the N-version programming robust to variations in few-shot example quality?}}

 \begin{table}[t]
\small
\centering
\caption{\revA{Sensitivity analysis of example quality with Qwen3-Coder-30B-A3B on the BIRD-Dev dataset.}}
\label{tab:icl_sensitivity}
\setlength{\tabcolsep}{4pt}
\renewcommand{\arraystretch}{1.05}
\begin{tabular}{llccc}
\toprule
\textbf{Generator} & \textbf{Retrieval Strategy} & \textbf{SQL Jaccard} & \textbf{EX (\%)} & \textbf{UB-EX (\%)} \\
\midrule
\multirow{2}{*}{Single ICL-based}
& Least Relevant & 0.44 & 70.1 & 75.8 \\
& Most Relevant  & 0.58 & 70.9 & 78.3 \\
\midrule
\multirow{2}{*}{Hybrid (ICL+D\&C+Skl)}
& Least Relevant & 0.44 & 71.5 & 80.2 \\
& Most Relevant  & 0.58 & 71.7 & 81.1 \\
\midrule
\multirow{2}{*}{\textbf{DeepEye-SQL (Full Pipeline)}}
& Least Relevant & 0.44 & 73.4 & 80.8 \\
& Most Relevant  & 0.58 & \textbf{73.5} & \textbf{81.6} \\
\bottomrule
\end{tabular}
\end{table}

\revA{To answer this, we explicitly tested the system's sensitivity by manipulating the retrieval strategy. Instead of the standard approach that selects the top-$K$ examples with the highest similarity (``Most Relevant''), we deliberately retrieved the $K$ examples with the lowest similarity (``Least Relevant''). 
We quantified the quality of these examples using the Jaccard similarity between the example SQL and the ground truth (following DAIL-SQL~\cite{dail-sql}). As shown in Table~\ref{tab:icl_sensitivity}, the ``Least Relevant'' set indeed exhibits significantly lower quality (0.44 vs. 0.58).
Under this challenging condition, the performance of the standalone ICL generator drops noticeably (70.9\% $\to$ 70.1\%). However, our Hybrid Generator dampens this drop (71.7\% vs. 71.5\%), and the full \model demonstrates remarkable stability (73.5\% vs. 73.4\%), verifying that our SE-inspired orchestration effectively masks the volatility of retrieval quality.}

\subsection{Confidence-aware SQL Selection Analysis}

\begin{figure}[t!]
    \centering
    \includegraphics[width=0.9\columnwidth]{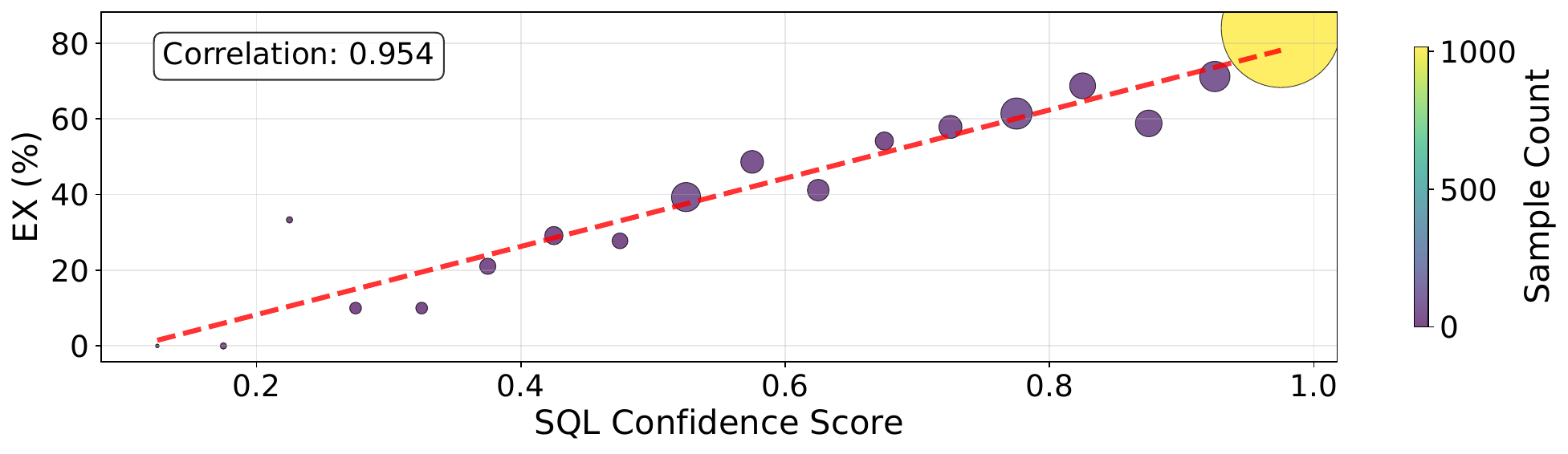}
    \caption{Execution accuracy \vs SQL confidence on BIRD-Dev dataset with Qwen3-Coder-30B-A3B model.}
    \label{fig:ex-vs-confidence}
\end{figure}


\begin{table}[t]
\small
\centering
\caption{Performance comparison of different SQL selection methods on the BIRD-Dev and Spider datasets.}
\label{tab:sql-selection-improvement}
\begin{tabular}{lcc}
\toprule
\textbf{Selection Method} & \textbf{BIRD-EX (\%)} & \textbf{Spider-EX (\%)} \\
\midrule
\multicolumn{3}{c}{\textit{\textbf{Gemma3-27B-Instruct}}} \\
\midrule
Consistency-based voting         & 70.1 & 88.3 \\
\textbf{Confidence-aware selection} & \textbf{71.1} ($\uparrow$1.0) & \textbf{88.9} ($\uparrow$0.6) \\
\midrule
\multicolumn{3}{c}{\textit{\textbf{Qwen2.5-Coder-32B-Instruct}}} \\
\midrule
Consistency-based voting         & 70.2 & 88.2 \\
\textbf{Confidence-aware selection} & \textbf{70.9} ($\uparrow$0.7) & \textbf{88.7} ($\uparrow$0.5) \\
\midrule
\multicolumn{3}{c}{\textit{\textbf{Qwen3-Coder-30B-A3B-Instruct}}} \\
\midrule
Consistency-based voting         & 72.7 & 89.7 \\
\textbf{Confidence-aware selection} & \textbf{73.5} ($\uparrow$0.8) & \textbf{89.8} ($\uparrow$0.1) \\
\bottomrule
\end{tabular}
\end{table}

\begin{figure}[t!]
    \centering
    \includegraphics[width=.9\columnwidth]{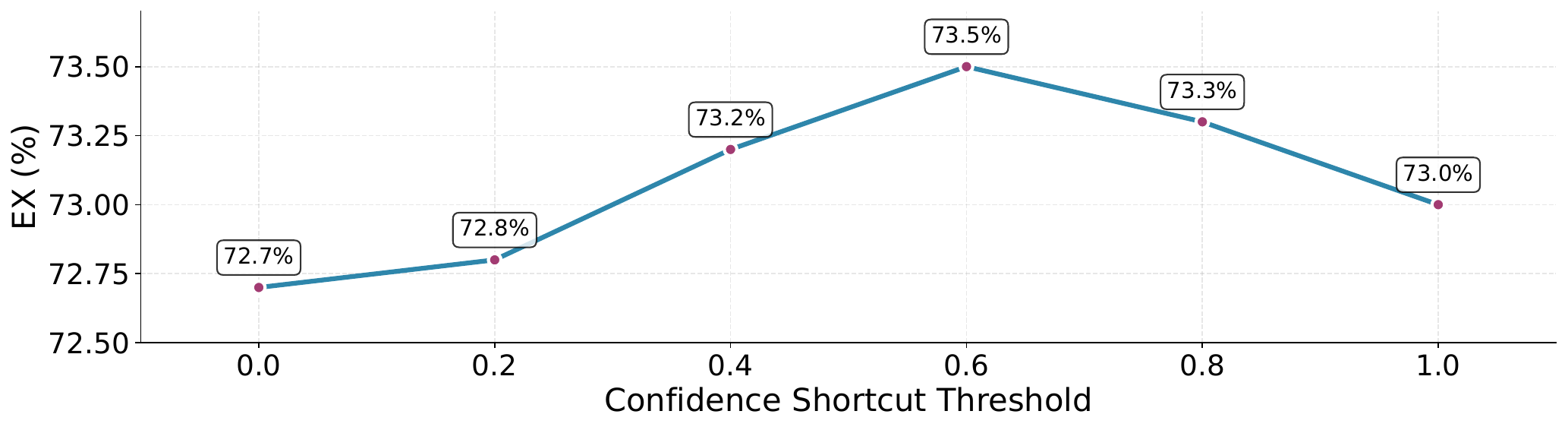}
    \caption{Execution accuracy \vs Confidence Shorcut Threshold on BIRD-Dev dataset with Qwen3-Coder-30B-A3B model.}
    \label{fig:threshold-sensitivity}
\end{figure}

\stitle{RQ6: Effectiveness of Confidence-aware Selection.}

We first validate that voting reliability strongly correlates with consensus. As shown in Figure~\ref{fig:ex-vs-confidence}, we observe a high Pearson correlation of \textbf{0.954} between voting confidence and execution accuracy on BIRD-Dev. This confirms that while high-confidence votes are reliable (justifying our efficient ``shortcut''), low-confidence scenarios require the deeper re-evaluation our method provides.
Comparative results in Table~\ref{tab:sql-selection-improvement} demonstrate that our approach consistently outperforms traditional Consistency-based Voting~\cite{alpha-sql, dail-sql}. For instance, with \texttt{Gemma3-27B}, we achieve a 1.0\% gain (70.1\% $\to$ 71.1\%) on BIRD, proving that selectively reviewing ambiguous cases yields superior robustness over blind majority voting.



\stitle{RQ7: Sensitivity to Confidence Threshold ($\theta_{conf}$).}

We evaluated the robustness of the threshold $\theta_{conf}$, which balances efficiency against rigorous review. Figure~\ref{fig:threshold-sensitivity} shows that performance remains highly stable across a wide range: accuracy stays above 73.0\% for $\theta_{conf} \in [0.4, 1.0]$, peaking at 73.5\% when $\theta_{conf}=0.6$. This indicates that our framework is not sensitive to specific hyperparameter choices.

\subsection{Ablation Study}


\begin{table}[t]
\small
\centering
\caption{Ablation study of \model with Qwen3-Coder-30B-A3B on the BIRD-Dev dataset.}
\label{tab:ablation-study}
\setlength{\tabcolsep}{4pt}
\renewcommand{\arraystretch}{1.04}
\begin{tabular}{lcc}
\toprule
\textbf{Variant} & \textbf{EX (\%)} & \textbf{$\Delta$ (\%)} \\
\midrule
\textbf{\model} & \textbf{73.5} & -- \\
\midrule
w/o Semantic Value Retrieval       & 71.4 & -2.1 \\
w/o Robust Schema Linking          & 71.8 & -1.7 \\
w/o Skeleton-based Generation      & 72.2 & -1.3 \\
w/o ICL-based Generation           & 71.0 & -2.5 \\
w/o D\&C-based Generation          & 72.3 & -1.2 \\
w/o Tool-Chain Testing \& Revision & 71.4 & -2.1 \\
w/o Confidence-aware Selection     & 72.7 & -0.8 \\
\bottomrule
\end{tabular}
\end{table}

\stitle{RQ8: What is the contribution of each key component to the overall performance of the \model framework?}

To understand the impact of each module, we conducted an ablation study on the BIRD-Dev dataset by progressively removing one component at a time. The results are detailed in Table~\ref{tab:ablation-study}.

The primary conclusion is that every component makes a positive and integral contribution, as removing any single module degrades performance. The \textit{ICL-based SQL Generation} is the most critical individual module, with its removal causing the largest performance drop of \textbf{2.5\%}. \textit{Semantic Value Retrieval} and \textit{SQL Unit Testing and Revision via Tool-Chain} are also highly impactful, each accounting for a \textbf{2.1\%} gain. This highlights the necessity of grounding the LLM in real data and externalizing the debugging process. The significant contributions from the SQL generation modules and the selection mechanism confirm the value of our N-version programming and confidence-aware selection strategies. Overall, the results confirm that the high performance of \model is not due to any single component, but rather the synergistic collaboration of all modules in its carefully designed pipeline.

\subsection{Efficiency Analysis}
\label{subsec:efficiency}

\begin{table}[t]
\small
\centering
\caption{\revC{Efficiency comparison on the BIRD-Dev dataset. \model achieves orders-of-magnitude lower latency than search-based (Alpha-SQL) and linear-chain (CHESS) baselines while maintaining higher accuracy.}}
\label{tab:efficiency}
\setlength{\tabcolsep}{4pt}
\renewcommand{\arraystretch}{1.03}
\begin{tabular}{lcccc}
\toprule
\multirow{2}{*}{\textbf{Method}} 
& \textbf{Avg. Input} 
& \textbf{Avg. Output} 
& \textbf{Avg. Latency} 
& \multirow{2}{*}{\textbf{EX (\%)}} \\
& \textbf{Tokens (K)} 
& \textbf{Tokens (K)} 
& \textbf{(s/query)} 
& \\
\midrule
CHESS (IR, SS, CS)      & 327.02 & 27.83 & 284.4 & 67.9 \\
Alpha-SQL               & 138.03 & 72.21 & 377.1 & 71.2 \\
\midrule
\textbf{\model}         & \textbf{23.21} & \textbf{23.16} & \textbf{17.8} & \textbf{73.5} \\
\quad Semantic Value Retrieval   & 0.67  & 0.03  & -- & -- \\
\quad Robust Schema Linking      & 13.91 & 6.33  & -- & -- \\
\quad N-version Generation       & 5.28  & 11.38 & -- & -- \\
\quad Tool-Chain Verification    & 3.16  & 5.41  & -- & -- \\
\quad Confidence-aware Selection & 0.19  & 0.01  & -- & -- \\
\bottomrule
\end{tabular}
\end{table}

\begin{figure}[t!]
    \centering
    \includegraphics[width=0.8\textwidth]{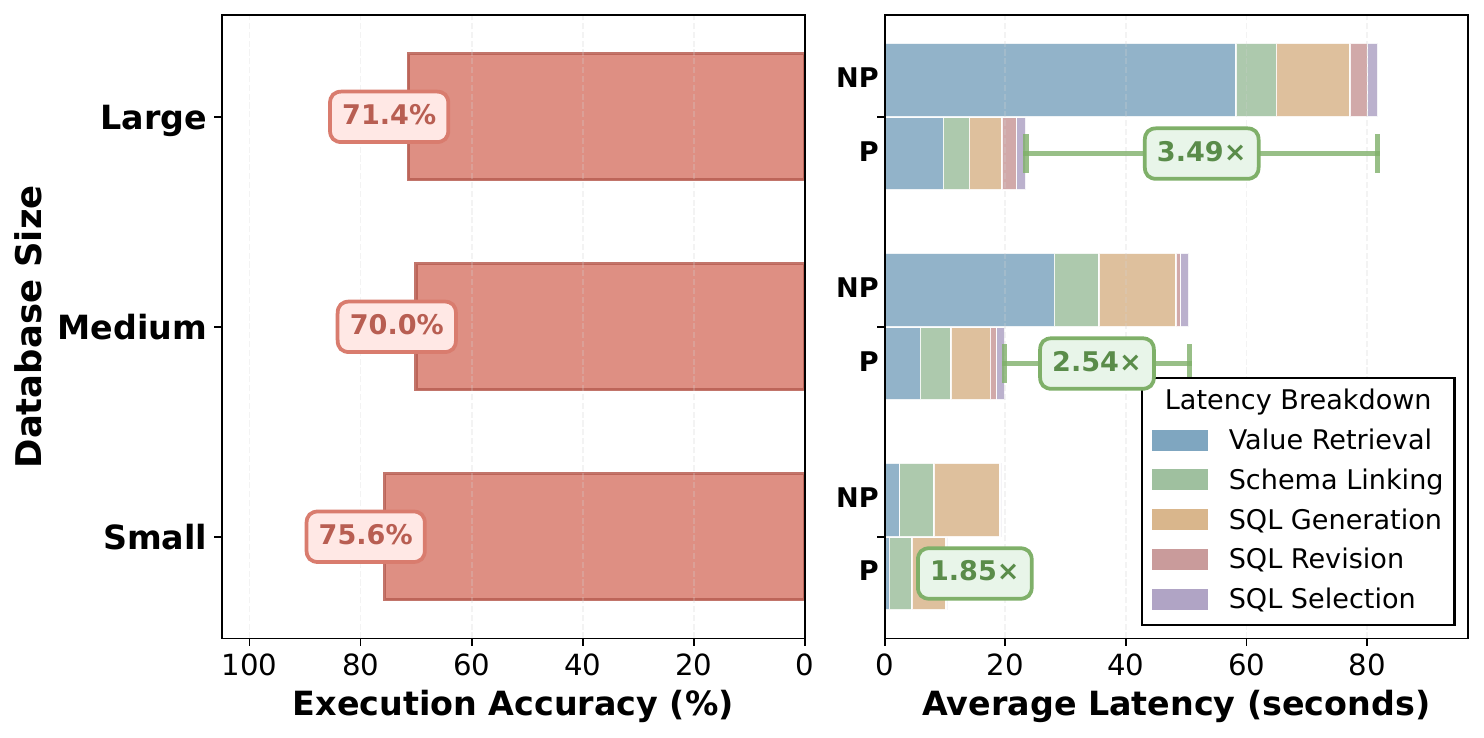}
    \caption{\revC{System Scalability Analysis on BIRD-Dev. (Left) Performance Robustness: We categorize databases into Small, Medium, and Large based on file size percentiles. \model maintains consistent execution accuracy even on Large schemas. (Right) Parallel Execution Efficiency: Comparison between Parallel (P) and Non-Parallel (NP) execution modes. The parallel pipeline significantly reduces latency, achieving a \textbf{3.49$\times$} speedup on Large databases by overlapping I/O-bound stages with Generation.}}
    \label{fig:scalability}
\end{figure}




\stitle{RQ9: How does \model perform in terms of efficiency and scalability compared to SOTA baselines?}

\revC{To answer this, we conducted a comprehensive evaluation covering token cost, end-to-end latency, and system scalability.}

\noindent\revC{\underline{\textbf{(i) Cost Efficiency (Token Consumption).}}}
\revC{As shown in Table~\ref{tab:efficiency}, \model demonstrates superior token efficiency. Compared to Alpha-SQL~\cite{alpha-sql} and CHESS~\cite{chess}, our framework reduces input token consumption by nearly \textbf{6$\times$} (138.03K vs. 23.21K) and \textbf{14$\times$} (327.02K vs. 23.21K), respectively, while achieving higher accuracy.}

\noindent\revC{\underline{\textbf{(ii) Time Efficiency (Latency).}}}
\revC{Beyond cost, time-to-result is critical for real-world deployment. Table~\ref{tab:efficiency} reveals that \model operates with an average latency of \textbf{17.8s} per query, which is orders of magnitude faster than search-centric baselines like Alpha-SQL (377.1s, $\sim$21$\times$ slower) and linear chains like CHESS (284.4s, $\sim$16$\times$ slower).
The high latency of Alpha-SQL stems from its sequential Monte Carlo Tree Search (MCTS), which inherently blocks parallelization. In contrast, \model's \textit{N-Version Programming} and \textit{Tool-Chain} are designed for parallel and conditional execution, maximizing throughput.}

\noindent\revC{\underline{\textbf{(iii) Scalability Analysis.}}}
\revC{To further assess robustness, Figure~\ref{fig:scalability} breaks down performance by database complexity.}
\begin{itemize}[leftmargin=*, topsep=2pt, itemsep=0pt]
    \item \revC{\textbf{Robust Accuracy (Figure~\ref{fig:scalability} Left):} \model maintains high execution accuracy across Small, Medium, and Large schemas, showing no significant degradation even as complexity increases.}
    \item \revC{\textbf{Parallel Acceleration (Figure~\ref{fig:scalability} Right):} The benefit of our parallel architecture scales with data size. On Large databases, enabling parallel execution yields a \textbf{3.49$\times$ speedup} (reducing latency from $\sim$80s to $\sim$23s) by effectively overlapping I/O-bound stages with Generation.}
\end{itemize}

\begin{figure}[t!]
    \centering
    \includegraphics[width=0.9\textwidth]{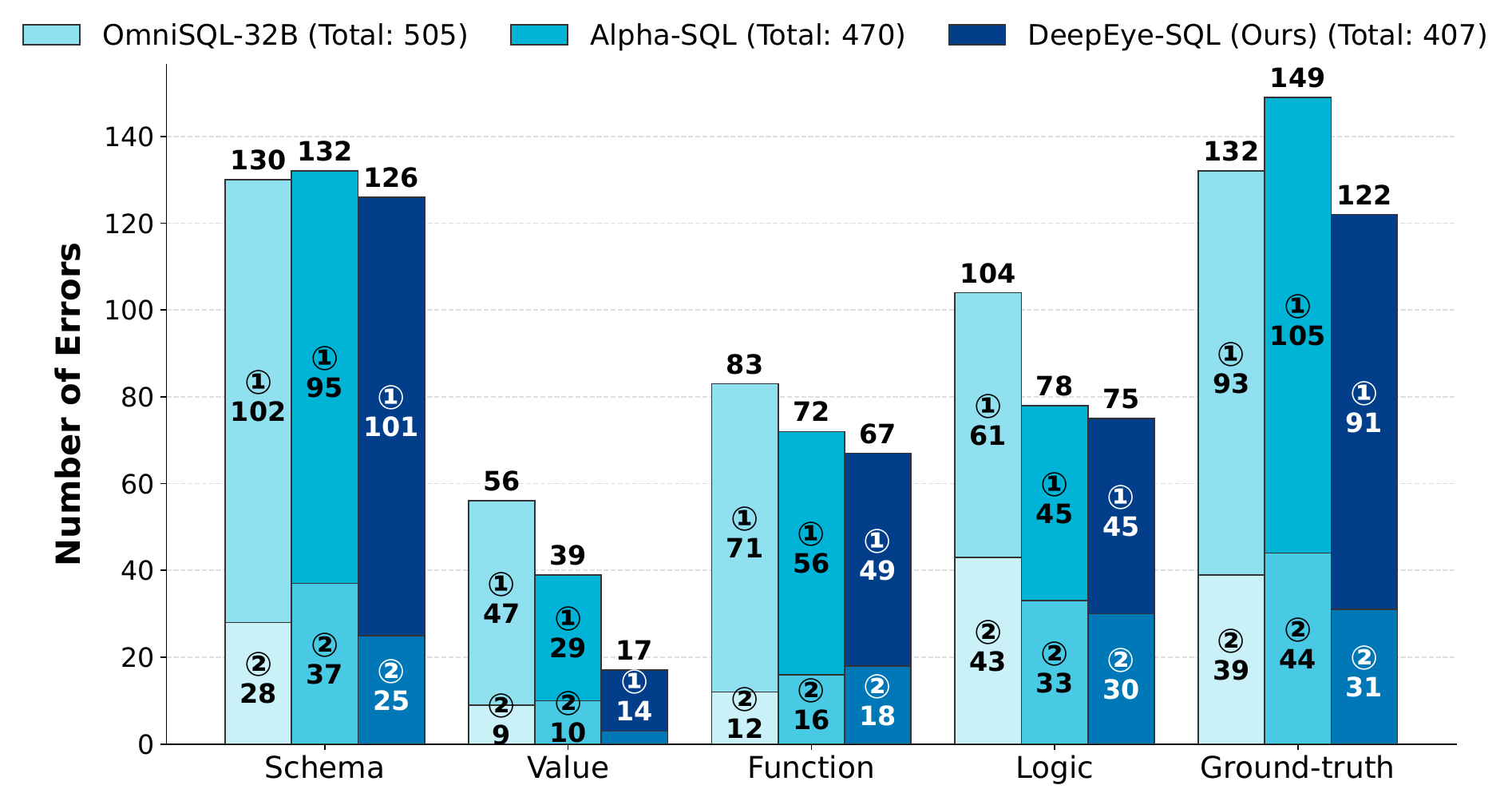}
    \caption{\revC{Comparative Error Analysis on BIRD-Dev.}}
    \label{fig:error_analysis}
\end{figure}

\subsection{\revC{Error Analysis}}
\label{subsec:error_analysis}
\stitle{\revC{RQ10: What are the dominant failure modes of \model, and how does its error distribution compare to SOTA baselines?}}

\revC{
To gain deeper insights into the limitations of \model, we performed a manual inspection of failure cases on the BIRD-Dev set. We compared \model with OmniSQL~\cite{omnisql} and Alpha-SQL~\cite{alpha-sql} using a fine-grained taxonomy covering five error dimensions, each divided into specific subclasses:
\textbf{Schema} (\ding{172} Selection, \ding{173} Join), 
\textbf{Value} (\ding{172} Mapping, \ding{173} Format), 
\textbf{Function} (\ding{172} Aggregation, \ding{173} Calculation), 
\textbf{Logic} (\ding{172} Filter, \ding{173} Presentation), 
and \textbf{Ground-truth} (\ding{172} Gold Error, \ding{173} Ambiguity).
}

\revC{
Figure~\ref{fig:error_analysis} illustrates the comparative distribution of these errors.
\model demonstrates a substantial advantage in handling \textit{Value-related} errors (only 17 instances), significantly outperforming baselines (56 and 39). This attributes to our \textit{Semantic Value Retrieval} strategy which effectively resolves both \ding{172} \textit{Value Mapping} and \ding{173} \textit{Data Format} issues.
Furthermore, the lower error rate in \textbf{Logic} and \textbf{Function} categories (67 and 75 respectively) validates the synergy of our \textit{N-version SQL Generation} and \textit{SQL Unit Testing \& Revision} modules.
However, \textbf{Schema Linking} remains the most challenging category for all models (126 errors for \model). Failures here are dominated by incorrect \ding{172} \textit{Column Selection} and \ding{173} \textit{Join Paths}, typically triggered by complex multi-hop requirements or obscure column names.
We also observed that roughly 30\% of the reported failures (122 cases) fall into the \textbf{Ground-truth} category, stemming from ambiguity in questions or errors in the gold SQLs, indicating the ceiling imposed by dataset quality.
}

\section{Related Work}
\label{sec:pre}

\stitle{Text-to-SQL Solutions.}
The task of translating natural language into executable SQL queries has evolved significantly~\cite{nl2sql-survey}. 
The recent emergence of Large Language Models (LLMs) has marked a new paradigm. Current research on applying LLMs to Text-to-SQL is largely divided into two categories: fine-tuning and prompting-based methods. Fine-tuning methods adapt open-source models like Code Llama or Qwen for the Text-to-SQL task by training them on large corpora of question-SQL pairs~\cite{codes, sense, omnisql, base-sql, xiyan-sql}. This approach can yield highly efficient and specialized models but may exhibit limited generalization to out-of-domain or highly complex scenarios.
In contrast, prompting-based methods leverage the powerful in-context learning and reasoning capabilities of very large, often proprietary, models like GPT-4o~\cite{gpt-4o} and Gemini~\cite{gemini} without requiring model training. To manage the complexity of the task, these methods typically decompose the process into a multi-stage pipeline, including sub-tasks like schema linking, SQL generation, and refinement~\cite{din-sql, supersql, chase-sql, alpha-sql, rewriter, elliesql}. While powerful, these frameworks often rely on a single generation path and the fallible self-correction abilities of the LLM, which can compromise robustness. Our work, \model, belongs to the prompting-based category but differentiates itself by explicitly adopting a systematic framework inspired by software engineering.

\stitle{Software Engineering.}
The challenge of building robust and reliable LLM-based systems often mirrors the complexities of traditional software development. In response, software engineering has established a set of core principles to manage complexity and ensure product quality. The \textit{Software Development Life Cycle (SDLC)}~\cite{sdlc} provides a foundational, systematic process for software creation, typically involving phases such as requirements analysis, implementation, testing, and deployment. To deconstruct existing systems and inform new designs, practitioners often employ \textit{Reverse Engineering}~\cite{reverse-engineering}, a process of analyzing a finished product to deduce its underlying specifications.
To enhance system reliability, fault tolerance techniques are critical. A notable example is \textit{N-Version Programming}~\cite{n-version-programming}, where multiple, independently developed versions of a component are executed in parallel, and their results are adjudicated to mask faults and increase the likelihood of a correct outcome. For quality assurance, \textit{Unit Testing}~\cite{unit-testing} is a fundamental practice. It involves the granular testing of individual software components or ``units'' in isolation to verify that each part functions correctly according to its design specifications. This bottom-up approach is crucial for identifying and rectifying defects early in the development process. Finally, to govern the release process, a \textit{Quality Gate}~\cite{quality-gate} acts as a final checkpoint, enforcing a set of predefined criteria to determine whether a software artifact meets the required quality standard for deployment.
%
Although central to traditional software engineering, these principles have seen limited systematic adoption in LLM-based pipelines. We bring them to Text-to-SQL to enable a structured, verifiable, and robust pipeline.

\section{Conclusion}
\label{sec:conclusion}
In this paper, we present \model, a verifiable SDLC-style workflow for Text-to-SQL that targets system-level reliability through end-to-end correctness.
\revB{Without any fine-tuning, \model leverages open-source MoE LLMs (about 30B total parameters with roughly 3B activated) and achieves 73.5\% execution accuracy on BIRD-Dev, 75.07\% on the official BIRD-Test leaderboard, and 89.8\% on Spider-Test, outperforming state-of-the-art approaches that rely on larger models or extensive training.}
Overall, these results suggest that principled orchestration, rather than fine-tuning or scaling up inference alone, is a promising path toward reliable Text-to-SQL.

\begin{acks}
This paper was supported by the NSF of China (62402409); Youth S\&T Talent Support Programme of Guangdong Provincial Association for Science and Technology (SKXRC2025461); the Young Talent Support Project of Guangzhou Association for Science and Technology (QT-2025-001); Guangdong Basic and Applied Basic Research Foundation (2023A1515110545); Guangzhou Basic and Applied Basic Research Foundation (2025A04J3935); and Guangzhou-HKUST(GZ) Joint Funding Program (2025A03J3714).
\end{acks}

\bibliographystyle{ACM-Reference-Format}
\bibliography{references}


\end{document}